\documentclass[apj]{emulateapj}       
\usepackage{apjfonts,epsfig,natbib}   

\newcommand{\hi}{\hbox{H~I}}          
\newcommand{\hii}{\hbox{H~II}}        

\newcommand{\otwo}{\hbox{[\ion{O}{2}] $\lambda$3727}}
\newcommand{\nethree}{[\ion{Ne}{3}] $\lambda$3869}
\newcommand{\stwoblue}{[\ion{S}{2}] $\lambda\lambda$4068,4076}
\newcommand{\hdelta}{\hbox{H$\delta$}}
\newcommand{\hgamma}{\hbox{H$\gamma$}}
\newcommand{\othreea}{[\ion{O}{3}] $\lambda$4363}
\newcommand{\hbeta}{\hbox{H$\beta$}}
\newcommand{\othree}{\hbox{[\ion{O}{3}] $\lambda\lambda$4959,5007}}

\newcommand{\othreec}{[\ion{O}{3}] $\lambda$5007}

\newcommand{\ooneb}{[\ion{O}{1}] $\lambda$6363}

\newcommand{\halpha}{\hbox{H$\alpha$}}
\newcommand{\ntwob}{[\ion{N}{2}] $\lambda$6583}

\newcommand{\stwo}{[\ion{S}{2}] $\lambda\lambda$6716,6731}

\newcommand{\otwored}{[\ion{O}{2}] $\lambda\lambda$7320,7330}

\journalinfo{ApJ, accepted (15 Dec 2005)}        

\shortauthors{Lee, Skillman, \& Venn}
\shorttitle{Nebular and Stellar Oxygen Abundances in NGC 6822}
\slugcomment{
Full version with color figures at {\tt http://www.astro.umn.edu/$\sim$hlee}
}

\begin{document}

\title{
The Spatial Homogeneity of Nebular and Stellar
Oxygen Abundances \\ in the Local Group Dwarf Irregular Galaxy 
NGC~6822$\,$\altaffilmark{1}
}

\author{
Henry Lee$\,$\altaffilmark{2}, 
Evan D. Skillman$\,$\altaffilmark{2}, and
Kim A. Venn$\,$\altaffilmark{2,3}
}

\altaffiltext{1}{
Based on EFOSC2 observations collected at the European Southern
Observatory, Chile: proposal \#71.B-0549(A). 
}
\altaffiltext{2}{
Department of Astronomy, University of Minnesota,
116 Church St. SE, Minneapolis, MN 55455;
{\tt hlee@astro.umn.edu, skillman@astro.umn.edu}
}
\altaffiltext{3}{
Department of Physics \& Astronomy, Macalester College,
1600 Grand Avenue, Saint Paul, MN 55105;
{\tt venn@macalester.edu}
}

\begin{abstract}			
To test the existence of a possible radial gradient in oxygen
abundances within the Local Group dwarf irregular galaxy NGC~6822, 
we have obtained optical spectra of 19 nebulae with the EFOSC2
spectrograph on the 3.6-m telescope at ESO La Silla.
The extent of the measured nebulae spans galactocentric radii in the
range between 0.05 kpc and 2 kpc (over four exponential scale lengths).
In five \hii\ regions (Hubble~I, Hubble~V, K$\alpha$, K$\beta$,
KD~28e), the temperature-sensitive \othreea\ emission line was
detected, and direct oxygen abundances were derived.
Oxygen abundances for the remaining \hii\ regions were derived using
bright-line methods.
The oxygen abundances for three A-type supergiant stars are slightly
higher than nebular values at comparable radii.
Linear least-square fits to various subsets of abundance data 
were obtained.
When all of the measured nebulae are included, no clear signature is
found for an abundance gradient.
%
%
A fit to only newly observed \hii\ regions with \othreea\ detections
yields an oxygen abundance gradient of $-0.14 \pm 0.07$ dex~kpc$^{-1}$.
The gradient becomes slightly more significant ($-0.16 \pm 0.05$ 
dex~kpc$^{-1}$) when three additional \hii\ regions with \othreea\
measurements from the literature are added. 
%
%
Assuming no abundance gradient, we derive a mean nebular oxygen
abundance 12$+$log(O/H) = $8.11 \pm 0.10$ from \othreea\ detections in
the five \hii\ regions from our present data; 
this mean value corresponds to [O/H] = $-0.55$.
\end{abstract}

\keywords{
galaxies: abundances --- 
galaxies: dwarf --- 
galaxies: evolution --- 
galaxies: individual (NGC~6822) ---
galaxies: irregular
}

\section{Introduction}			

The evolution of element abundances with time provides important
clues to the history of chemical enrichment and star formation
in galaxies.
Abundances in \hii\ regions provide information about the most recent
period of metals enrichment in the interstellar medium, now
illuminated by recently-formed O, B-type stars.
Dwarf irregular galaxies contain few metals and are thought to be
chemically homogeneous.
These low-mass, metals-poor, gas-rich, and star-forming systems
provide unique venues to examine detailed processes of star formation
within environments which may be related to star-forming systems 
observed at early times.
Thus far, no nearby ($D \approx 5$ Mpc) dwarf irregular galaxy has ever
exhibited a significant radial abundance gradient
(i.e., \citealp{ks96,ks97,ls04,lsv05}). 
A possible reason is that the most recent burst of star formation
has expelled the most recent synthesis of metals into the hot phase of
the interstellar medium, and that these metals have yet to cool
and ``rain'' back down onto the cooler phases of the interstellar
medium (e.g., \citealp{tt96,ks97}).
It would indeed be very interesting to find an example of a dwarf
galaxy with localized oxygen enrichment.
Recent developments in high-efficiency spectrographs on 8- and 10-m
telescopes have made possible spectroscopy of bright blue A-type
supergiants in nearby dwarf irregular galaxies 
(e.g., \citealp{venn03,kaufer04}). 
These young hot massive stars allow for the simultaneous measurements
of present-day $\alpha$- and iron-group elements.
These measurements also allow for the direct comparison of stellar
$\alpha$-element abundances with nebular measurements, as massive
stars and nebulae are similar in age and have similar formation sites.

NGC~6822 is the nearest gas-rich dwarf irregular galaxy in the Local
Group \citep{mateo98}.
Basic properties are listed in Table~\ref{table_n6822}; see
also \cite{vdb00}. 
Studies of the stellar populations and the star formation history of
NGC~6822 are described by 
\cite{gallart96a,gallart96b,gallart96c}, \cite{cb98},
\cite{hutchings99}, \cite{tolstoy01}, \cite{wyder01}, and
\cite{clementini03}.
A large \hi\ halo extending much farther out than the optical extent
was discovered by \cite{roberts72}, and confirmed by \cite{debw00} 
and \cite{komiyama03}.
A population of young, blue stars was found, whose spatial distribution 
is farther than the accepted optical radius and is very similar to
that of the \hi\ extent \citep{battinelli03,debw03,komiyama03}.
With the characterization of the spatial and color-magnitude
distributions of red giant and asymptotic giant branch stars across
the entire galaxy in the near-infrared, \cite{ch05} found a
metallicity spread of 1.56~dex from the ratio of carbon- to
M-type stars. 
\cite{leisy05} have recently reported 13 new candidate planetary
nebulae (PNe), bringing to 17 total PNe in all.
\cite{musch99} obtained spectra of three B-type supergiant
stars, and reported a mean iron abundance of 
[Fe/H] = $-0.5 \pm 0.2$$\,$\footnote{
We use the notation: [X/Y] = log(X/Y) $-$ log(X/Y)$_{\odot}$.
}.
\cite{venn01} reported for the first time oxygen abundances for two
A-type supergiant stars in NGC~6822, and showed that the mean stellar
oxygen abundance was higher than the known \hii\ region nebular
abundances by at least 0.1~dex. 
Because their stars were at low galactocentric radii, the authors
suggested the possibility of a radial gradient in oxygen abundance.

The recent measurements of oxygen abundances in A-type supergiant
stars have motivated the reevaluation of published nebular oxygen
abundances in NGC~6822. 
Spectra of and abundances for the brightest \hii\ regions in NGC~6822
were reported by \cite{ps70}, 
\cite{alloin74}\footnote{ 
\cite{alloin74} did not derive an oxygen abundance as
their \othree\ emission lines were saturated and not measured.
},
\cite{smith75}, \cite{lequeux79}, \cite{talent80}, and \cite{pes80}.
However, all of these spectra were obtained with inherently nonlinear
detectors, and in many cases the character of the nonlinearities were
not understood until well after publication 
(e.g., \citealp{jenkins87}); so, subsequent corrections for
nonlinearity were not possible. 
\cite{stm89} obtained CCD spectroscopy to measure \othreea\ in
Hubble~V, and derived an oxygen abundance of 12$+$log(O/H) = 8.20.
In two planetary nebulae, \cite{rm95} derived oxygen abundances
(12$+$log(O/H) = 8.01, 8.10) in agreement with
published \hii\ region oxygen abundances.
\cite{miller96} remeasured Hubble~V and Hubble~X, and derived oxygen
abundances 12$+$log(O/H) = 8.32 and 8.36, respectively.
In their program on open clusters, \cite{chandar00} also obtained
spectra for a few nebulae, and derived smaller nebular oxygen
abundances than expected.
We have reanalyzed a number of their \hii\ region spectra, which we
discuss in Sect.~\ref{sec_discuss}.
\cite{hgom01} also obtained spectra of Hubble~V and Hubble~X, and
while there were no large differences in oxygen abundances between
the two \hii\ regions, they claimed small-scale abundances variations 
on $<$ 10~pc length scales.
\cite{josh02} derived sulfur abundances from {\em ISO\/} measurements
of [\ion{S}{3}] and [\ion{S}{4}] emission lines in the mid-infrared,
and showed that the S$^{+3}$ ion is the largest contributor to the
total sulfur abundance in extragalactic \hii\ regions.
With VLT data, \cite{peimbert05} derived recombination-line
abundances for the \hii\ region Hubble~V, and showed that their
derived oxygen abundances were in better agreement with the 
stellar oxygen abundances.
Some of these results are discussed further 
in Sec.~\ref{sec_recentwork} below. 

This is the second of two papers of our study examining the spatial
homogeneity of oxygen abundances in Local Group dwarf irregular
galaxies; WLM was discussed previously in \cite{lsv05}. 
The main goal here was to obtain a homogeneous set of nebular spectra
for \hii\ regions in NGC~6822 over a large range in 
galactocentric radii.
The outline of this paper is as follows.
Descriptions of the observations, reductions, measurements
and analysis are presented in Sect.~\ref{sec_obsmeas}.
Element abundances and abundance ratios are described 
in Sect.~\ref{sec_abund}. 
In Sect.~\ref{sec_discuss}, we compare present results 
with recent studies, and examine the presence of spatial
inhomogeneities in oxygen abundances.
A summary is given in Sect.~\ref{sec_concl}.
For the present discussion, we use the notation
[O/H] = log(O/H) $-$ log(O/H)$_{\odot}$, where the solar value
of the oxygen abundance is 12$+$log(O/H) = 8.66
\citep{asplund04,melendez04}.

\section{Observations and Measurements}	
\label{sec_obsmeas}		

\subsection{Observations and Reductions}

Long-slit spectroscopic observations of nebulae in 
NGC~6822 were carried out on 2003 Aug. 26--28 and 31 (UT) with the 
ESO Faint Object Spectrograph and Camera (EFOSC2) instrument on the
3.6-m telescope at ESO La Silla Observatory.
``Blue'' spectra with a smaller wavelength range and low-dispersion
spectra with larger wavelength coverage into the red were obtained
with gratings 7 and 11, respectively.
Details of the instrumentation employed and the log of observations
are listed in Tables~\ref{table_obsprops} and \ref{table_obslog},
respectively.
Observations were obtained during new moon phase.
Two-minute \halpha\ acquisition images were obtained in
order to set an optimal position of the slit.
Typically, the slit angle was set to obtain spectra for more than one
\hii\ region.
Average departures of the slit position angle from the parallactic
angle are listed in Table~\ref{table_obslog}.
Nineteen nebulae for which spectra were obtained are listed
in Table~\ref{table_obslog} and are identified in
Figs.~\ref{fig_n6822_gxy} to \ref{fig_n6822_ctr}.
Identifications for the nebulae follow from the \halpha\ imaging
by \cite{kd82} and \cite{hkl88}, and are matched with the \halpha\
image from the Local Group Survey \citep{massey_lgs}\footnote{
More about the Local Group Survey is found at 
\url{{\tt http://www.lowell.edu/users/massey/lgsurvey.html}}.
}.
The well-studied bright \hii\ region Hubble~V
(e.g., \citealp{smith75,talent80,pes80,stm89,miller96,hgom01}) was
also observed to check the reliability of our measurements for
oxygen abundances. 

Data reductions were carried out in the standard manner using 
IRAF\footnote{
IRAF is distributed by the National Optical Astronomical
Observatories, which are operated by the Associated Universities for
Research in Astronomy, Inc., under cooperative agreement with the
National Science Foundation.}
routines.
Data obtained on a given night were reduced independently.
The raw two-dimensional images were trimmed and the bias level was
subtracted.
Dome flat exposures were used to remove pixel-to-pixel variations 
in response. 
Twilight flats were acquired at dusk each night to correct
for variations over larger spatial scales.
To correct for the ``slit function'' in the spatial direction, the
variation of illumination along the slit was taken into account
using dome and twilight flats. 
Cosmic rays were removed in the addition of multiple exposures
for a given \hii\ region.
Wavelength calibration was obtained using helium-argon (He-Ar) arc
lamp exposures taken throughout each night.
Exposures of standard stars Feige~110, G138$-$31, LTT~1788, LTT~7379,
and LTT~9491 were used for flux calibration.
The flux accuracy is listed in Table~\ref{table_obslog}.
Final one-dimensional spectra for each \hii\ region were obtained via
unweighted summed extractions.  

\subsection{Measurements and Analysis}	

Emission-line strengths were measured using software developed 
by M. L. McCall and L. Mundy; see \cite{lee01}
and \cite{lee03field,lee03virgo}.
In all, \othreea\ was detected in five \hii\ regions; these spectra
are shown in Figs.~\ref{fig_gr07spc} and \ref{fig_gr11spc}.
The corrections for reddening and underlying Balmer absorption
are described in \cite{ls04} and \cite{lsv05}.
For nebulae with blue spectra, 
observed flux $(F)$ and corrected intensity $(I)$ ratios are listed
in Tables~\ref{table_gr07data1} to \ref{table_gr07data2} inclusive.
Flux and intensity ratios for nebulae with low-dispersion spectra
are presented in Tables~\ref{table_gr11data1} to \ref{table_gr11data3}
inclusive.
In high signal-to-noise spectra, we derived the logarithmic reddening,
$c(\hbeta)$, from the error weighted average of values
for $F(\halpha)/F(\hbeta)$, $F(\hgamma)/F(\hbeta)$, and
$F(\hdelta)/F(\hbeta)$ ratios, while simultaneously solving for the
effects of underlying Balmer absorption with equivalent
width, EW$_{\rm abs}$; see Fig.~\ref{fig_monte} for \hii\ region
Hubble~V.
We assumed that EW$_{\rm abs}$ was the same for \halpha, \hbeta,
\hgamma, and \hdelta.
In the data tables, we have included the logarithmic reddening and the
equivalent width of the underlying Balmer absorption at \hbeta.
Where negative values were derived, the reddening was set to zero
in correcting line ratios and for abundance calculations.

\section{Nebular Abundances}
\label{sec_abund}

\subsection{\hii\ Regions}

Oxygen abundances in \hii\ regions were derived using three methods:
(1) the direct method (e.g., \citealp{dinerstein90,skillman98});
and the bright-line methods discussed by
(2) \cite{mcgaugh91}, which is based on photoionization models;  
and (3) \cite{pilyugin00}, which is purely empirical.
These methods are explained in detail in \cite{ls04} and \cite{lsv05};
we briefly summarize them here.

\subsubsection{Oxygen Abundances: \othreea\ Temperatures}
\label{sec_direct}

The ``direct'' conversion of emission-line intensities into ionic
abundances requires a reliable estimate of the electron temperature 
of the ionized gas.
We adopt a two-zone model for \hii\ regions, with a low- and a
high-ionization zone characterized by temperatures $T_e($O$^+)$ and
$T_e($O$^{+2})$, respectively. 
The temperature in the O$^{+2}$ zone is measured with
the emission-line ratio $I$(\othreec)/$I$(\othreea) 
\citep{osterbrock}.
The temperature in the O$^+$ zone is given by
\begin{equation}
t_e({\rm O}^+) = 0.7 \; t_e({\rm O}^{+2}) + 0.3,
\label{eqn_toplus}
\end{equation}
where $t_e = T_e/10^4$~K \citep{ctm86,garnett92}.
The total oxygen abundance by number is given by
O/H = O$^+$/H$^+$ $+$ O$^{+2}$/H$^+$.
For conditions found in typical \hii\ regions and those presented
here, very little oxygen in neutral form is expected, and in the
absence of He~II emission, the O$^{+3}$ contribution is considered
negligible.
For subsequent calculations of ionic abundances, we assume
the following electron temperatures for specific ions
\citep{garnett92,til95}: 
$t_e$(N$^+$) = $t_e$(O$^+$), 
$t_e$(Ne$^{+2}$) = $t_e$(O$^{+2}$), 
$t_e$(Ar$^{+2}$) = 0.83 $t_e$(O$^{+2}$) $+$ 0.17, and
$t_e$(Ar$^{+3}$) = $t_e$(O$^{+2}$).

Derived ionic and total abundances are listed in
Tables~\ref{table_abund1} and \ref{table_abund2}, which include
derived O$^+$ and O$^{+2}$ electron temperatures, O$^+$ and O$^{+2}$
ionic abundances, and the total oxygen abundances.
Direct \hii\ region oxygen abundances were derived for Hubble~I,
Hubble~V, K$\alpha$, K$\beta$, and KD~28e, and were in excellent
agreement with those derived from the method described by
\cite{scm03}.

Where \stwoblue\ and \otwored\ were detected in
the spectra for Hubble~V, K$\alpha$, and KD~28e, we used
the IRAF task {\tt temden} in the STSDAS {\tt nebular} package
\citep{sd95} to derive $T_e$(S$^+$) and $T_e$(O$^+$) from
$I$(\stwo)/$I$(\stwoblue) and $I$(\otwo)/$I$(\otwored),
respectively.
$T_e$(S$^+$) would be too low, if \stwoblue\ was faint
and barely detected.
$T_e$(O$^+$) would be too high, if the \otwored\ flux was too high,
which would arise from second-order contamination at wavelengths
below 3700~\AA. 
However, we did not use a second-order blocking filter in
our spectroscopy program.
Nevertheless, we list these temperatures in Tables~\ref{table_abund2}
and \ref{table_abund3}.

Electron densities were derived for \hii\ regions 
Hubble~V, K$\alpha$, and KD~28e from their low-dispersion spectra.
With their computed O$^{+2}$ temperatures, the electron densities are
63, 332, and 123 cm$^{-3}$, respectively.
Derived values for the electron density did not change significantly
with an assumed O$^{+2}$ temperature of $10^4$~K, or when O$^+$
temperatures are used in the computations.
Also, very little change in oxygen abundances for Hubble~V, K$\alpha$,
and KD~28e occurred when an electron density of 100 cm$^{-3}$ was
assumed. 

\subsubsection{Oxygen Abundances: Bright-Line Methods}
\label{sec_brightline}

In \hii\ regions without \othreea\ measurements, the bright-line
method has been used to derive oxygen abundances, as the latter
are given in terms of bright [O~II] and [O~III] emission lines,
i.e., $R_{23}$ = [$I$(\otwo) + $I$(\othree)]/$I$(\hbeta)
indicator suggested by \cite{pagel79} and discussed by
\cite{skillman89}. 
\cite{mcgaugh91} developed a grid of photoionization models and
suggested using $R_{23}$ and ionization factor,
$O_{32}$ = $I$([\ion{O}{3}])/$I$([\ion{O}{2}]),
to estimate the oxygen abundance\footnote{
Analytical expressions for the McGaugh calibration
can be found in \cite{chip99}.
}.
To break the degeneracy in the bright-line method,
we have used the $I$([\ion{N}{2}]/$I$([\ion{O}{2}] ratio 
(e.g., \citealp{mrs85,mcgaugh94,vanzee98,lee03field})
to choose between the ``upper branch'' (high
oxygen abundance) or the ``lower branch'' (low oxygen abundance).
In some instances, oxygen abundances with the McGaugh method could
not be computed, because the $R_{23}$ values were outside of the
effective range for the models.
\citet[Equation~4]{pilyugin00} proposed an empirical calibration at
low metallicity with fits of oxygen abundance against bright oxygen
lines.
\cite{scm03} have shown that while the Pilyugin calibration 
covers a larger range in $R_{23}$, the calibration applies mostly
to \hii\ regions with higher ionizations; see also the discussion
by \cite{vzh05}.

Oxygen abundances derived using the McGaugh and Pilyugin bright-line
calibrations are listed in Tables~\ref{table_abund1} and
\ref{table_abund2}.
For each \hii\ region, differences between direct and bright-line
abundances are shown as a function of $O_{32}$ and $R_{23}$
in Fig.~\ref{fig_oxydiff}.
The difference between the McGaugh and Pilyugin calibrations 
(indicated by asterisks) correlates with log $O_{32}$, 
which has been discussed by \cite{scm03}, \cite{lee03south}, and
\cite{ls04}.
We find that bright-line abundances with the McGaugh and the Pilyugin
calibrations are about $\pm 0.10$~dex and up to 0.20~dex larger,
respectively, compared to the corresponding direct abundances.
Generally, in the absence of \othreea, an estimate of the oxygen
abundance from the bright-line calibration is good to within
$\approx$ 0.2~dex.

\subsection{Supernova Remnants}
\label{sec_snr}

The emission-line ratio criterion\footnote{
This is hereafter referred as $I$([\ion{S}{2}])/$I$(\halpha).
}
$I$(\stwo)/$I$(\halpha)
$\ga 0.4$ has been used 
to distinguish supernova remnants (SNRs) from photoionized \hii\
regions and planetary nebulae at near-solar metallicities
(e.g., \citealp{ddb80,smith75,smith93,levenson95}). 
\cite{skillman85} found that strong [\ion{O}{1}] emission 
($I$([\ion{O}{1}])/$I$(\hbeta) $\ga 0.1$) indicates the presence of
shocks from supernova remnants; whereas in typical \hii\ regions,
there is very little oxygen in the form of neutral oxygen (e.g.,
\citealp{bpt81,vo87,stasinska90}).
We list examples of oxygen abundances in SNRs within nearby dwarf
galaxies.
\cite{miller95} used the shock models of \cite{dopita84} to 
estimate oxygen abundances for two shock-heated nebula in the M~81 group
dwarf galaxy Holmberg~IX, and found that the estimated mean oxygen
abundance (12$+$log(O/H) $\approx$ 8) was comparable to other dwarf
galaxies at similar total (optical) luminosity.
Applying the MAPPINGS code to spectrophotometric data, 
\cite{rd90} determined that the mean oxygen abundances for
SNRs in the Magellanic Clouds were 0.12 to 0.20 dex lower than the
mean oxygen abundances derived for the \hii\ regions. 
However, the errors associated with the abundance differences were
also consistent with zero difference between the SNRs and the
\hii\ regions.
We should note that the line ratios used to determine
abundances for supernova remnants have been calibrated to the
metallicity of the Milky Way, and that the line ratios are
themselves metallicity-dependent; 
i.e., $I$([\ion{S}{2}])/$I$(\halpha) decreases with decreasing
metallicity.

Hodge~12 has been long known as an SNR 
(e.g., \citealp{ddb80,dd83,kong04}).
\cite{smith75} obtained spectroscopy and found flux ratios 
[\ion{S}{2}]/\halpha\ = 0.5 and [\ion{O}{3}]/\hbeta\ = 0.9.
Our two-dimensional spectrum of Hodge~12 and the resulting extracted
one-dimensional spectrum are shown in Figs.~\ref{fig_snr2d}
and \ref{fig_snr}, respectively.
Our measurements (Table~\ref{table_gr11data1}) are in general
agreement with published line ratios: 
$I$(\othree)/$I$(\hbeta) $\simeq 0.8$, 	    
$I$(\ooneb)/$I$(\hbeta) $\simeq 0.2$,       
$I$(\ntwob)/$I$(\halpha) $\simeq 0.09$, and 
$I$([\ion{S}{2}])/$I$(\halpha) $\simeq 0.5$.	    
Using the grid of SNR models from \citeauthor{dopita84}
\citeyearpar[Fig.~8]{dopita84} and our $I$([\ion{S}{2}])/$I$(\halpha)
value, we estimate an oxygen abundance of 12$+$log(O/H)  
$\simeq$ $7.9 \pm 0.1$, although our $I$(\otwo)/$I$(\hbeta) and
$I$(\othree)/$I$(\hbeta) ratios are, respectively, higher and lower
than the ranges expressed in their models.
The error in the oxygen abundance is an estimate of the 
possible range of values obtained from the models.

There are possible SNR among the observed ionized nebulae.
The nebulae KD~20 (identified as C10 by \citealp{chandar00}) has an
$I$([\ion{S}{2}])/$I$(\halpha) ratio of 0.4, which is roughly four times
larger than the value for a typical \hii\ region in NGC~6822.
This excess [\ion{S}{2}] emission is likely the signature of the
presence of an SNR.
Note that the canonical ratio of 0.5 was established for solar
metallicities, and at lower metallicities, [\ion{S}{2}] emission
in a photoionized region is smaller; so,
$I$([\ion{S}{2}])/$I$(\halpha) ratios below one-half can 
still indicate the presence of shock excitation in metal-poor
galaxies \citep{skillman85}.
The nebula KD~12 (identified as C9 by \citealp{chandar00}),
although not measured here in the present work, has a similar
value of $I$([\ion{S}{2}])/$I$(\halpha), and is likely another SNR.
In the absence of measured [\ion{N}{2}], we adopt the oxygen abundance
for KD~12 measured by \cite{chandar00} as a lower limit.

\subsection{Element Ratios}

We briefly discuss argon-to-oxygen, nitrogen-to-oxygen,
and neon-to-oxygen ratios, which are listed in
Tables~\ref{table_abund1} and \ref{table_abund2}.
The following discussion can be referred to Fig.~\ref{fig_zratios}.

We derived nitrogen-to-oxygen ratios for three \hii\ regions with
\othreea\ and \ntwob\ measurements: Hubble~V, K$\alpha$, and KD~28e.
Although \ntwob\ is blended with \halpha\ in our low-dispersion
spectra, we used the IRAF {\tt splot} routine to
deblend and measure their line fluxes.
For metal-poor galaxies, it is assumed that 
N/O $\approx$ N$^+$/O$^+$ \citep{garnett90} and N$^+$/O$^+$ 
values were derived.
Nitrogen abundances were computed as N/H = ICF(N) $\times$ (N$^+$/H).
The ionization correction factor, ICF(N) = O/O$^+$, accounts for
missing ions. 
The resulting nitrogen-to-oxygen abundance ratios were found
to be the same as the N$^+$/O$^+$ values.
From the measurements of the three \hii\ regions, mean values are: 
N/O = $0.012 \pm 0.001$,	    
and log(N/O) = $-1.92 \pm 0.04$.    
The latter is significantly lower than the mean for metal-poor blue
compact dwarf galaxies (log(N/O) = $-1.46 \pm 0.14$ for
galaxies with 12$+$log(O/H) $\ga 7.6$; \citealp{it99}).
Such low values of N/O are not found at all within the sample of
blue compact dwarf galaxies by \cite{it99}, although some objects
with similar N/O ratios are found in the compilation from the
literature by \cite{ks96}.
\citeauthor{vanzee97a} \citeyearpar{vanzee97a,vanzee97b} have found
similarly low values of N/O in low surface brightness galaxies, 
although \cite{it99} have questioned the quality of these data.

\cite{stm89} proposed that the low N/O value in NGC~6822 could be
attributed to a recent burst of star formation, and can be explained
by the time delay between the release of newly synthesized oxygen from
massive stars and newly synthesized nitrogen from intermediate mass
stars. 
This was motivated in part by the discovery of a period of
star cluster formation about 100~Myr ago \citep{hodge80}.
\cite{gallart96c} used extensive color-magnitude diagram analysis of
NGC~6822 to determine the occurrence of an enhanced episode of star
formation between 100 and 200~Myr ago.
The hypothesis that relatively low N/O observed in NGC~6822 is real
and due to its history of star formation history appears to
be supported by earlier observations of low N/O values 
\citep{pes80,stm89} and by the low values in three
different nebulae from the present study.
However, we caution the reader that the present N/O values were
derived from low-dispersion spectra with \halpha\ and \ntwob\ blended
in a number of the spectra (see Figs.~\ref{fig_gr11spc} and \ref{fig_snr}).
The larger N/O values reported by \cite{peimbert05} (in better
agreement with the mean value for blue compact dwarf galaxies) were
derived from higher-dispersion spectra obtained with the VLT.
Due to the importance of the proper interpretation of the
N/O values with the recent history of star formation, the
N/O values should be confirmed.

Neon abundances are derived as Ne/H = ICF(Ne) $\times$ (Ne$^{+2}$/H$^+$).
The ionization correction factor for neon is ICF(Ne) = O/O$^{+2}$.
%
%
The mean is log(Ne/O) = $-0.59 \pm 0.12$, 
which is just in agreement with the mean value of $-0.72 \pm 0.06$ for
blue compact dwarf galaxies \citep{it99}\footnote{
Note that part of the difference is due to a 14\% difference in the
ratio of the \othreec\ and \nethree\ emissivities used by
\cite{it99} and that computed by the {\tt IONIC} task in the 
NEBULAR code of \cite{sd95}.
This 14\% difference, which translates into a 0.06 dex difference (in
the sense observed), is probably an indication of the minimum
systematic uncertainty in the atomic data which are used for
calculating nebular abundances (see \citealp{garnett04}).
}.
Since the $F$(\nethree)/$F$(\othreec) ratio is sensitive to the
reddening correction, and our reddening corrections are only based on
the $F$(\halpha)/$F$(\hbeta) ratio, we revisited this correction
method.
We have used only reddenings derived using $F$(\halpha)/$F$(\hbeta)
in spectra where \othreea\ was detected.
If the $F$(\halpha)/$F$(\hbeta) is too large,
$F$(\nethree)/$F$(\hbeta) may be overcorrected. 
In the spectra for \hii\ regions Hubble~V, K$\alpha$, K$\beta$,
and KD~28e, high-order unblended Balmer lines (H9, H10, and H11) were
detected. 
The intensity ratios with respect to \hbeta\ were found to be
consistent with expected values for \hii\ regions having electron
temperatures in the range between 11500 and 14000~K.
In the blue spectrum for \hii\ region Hubble~I, we did
not detect the higher-order Balmer lines as above.
The unblended corrected Balmer line closest to \nethree\ is \hdelta,
because H8 is blended with an adjacent helium line, and H$\epsilon$ is
blended with adjacent [\ion{Ne}{3}] and helium lines. 
We found that the resulting $I$(\hdelta)/$I$(\hbeta) and 
$I$(\hgamma)/$I$(\hbeta) ratios were consistent with expected Balmer
ratios at a temperature $T_e({\rm O}^{+2})$ of about 13000~K. 

Argon is more complex, because the dominant ion is not found in
just one zone.
Ar$^{+2}$ is likely to be found in an intermediate area between the
O$^+$ and O$^{+2}$ zones.
Argon abundances were derived using the prescription by \cite{til95}.
If only [\ion{Ar}{3}] is present,
the argon abundance is Ar/H = ICF(Ar) (Ar$^{+2}$/H$^+$), and
the ionization correction factor is given by 
ICF(Ar) = Ar/Ar$^{+2}$ = $[0.15 + x(2.39 - 2.64x)]^{-1}$,
where $x$ = O$^+$/O.
Where [\ion{Ar}{4}] is also present, the argon abundance is Ar/H = 
ICF(Ar) ((Ar$^{+2}$ + Ar$^{+3}$)/H$^+$), and
the ionization correction factor is given by 
ICF(Ar) = Ar/(Ar$^{+2}$ + Ar$^{+3}$) = 
$\{0.99 + x[0.091 + x(-1.14 + 0.077x)]\}^{-1}$.
%
%
Our mean value of log(Ar/O) = $-2.10 \pm 0.06$ is in agreement with
the average for metal-poor blue compact dwarf galaxies \citep{it99}.

\cite{moore04} has recently suggested that direct modeling of
photoionized nebulae should be used to infer elemental abundances with
accuracies similar to observations.
Abundances derived from model-based ionization correction factors were
shown to exceed the range of expected errors from the original data.

\section{Interstellar Medium and Stellar Oxygen Abundances in NGC~6822}
\label{sec_discuss}           

A compilation of measured oxygen abundances from \hii\ regions,
planetary nebulae, and stars from the present work and from the
literature is presented in Table~\ref{table_alloxy}.
To ascertain the possibility of spatial variations in oxygen
abundance across the galaxy, we first discuss how radial gradients
are derived, followed by a discussion of recent studies in the
literature. 
We examine our highest-quality results in the context of the
best data reported in the literature.
We will then discuss how the existence of a nonzero radial
abundance gradient and a zero gradient can be interpreted from
the data, keeping in mind the relative dispersion in oxygen abundances
used for the radial fits.

Deprojected galactocentric radii were derived similar to the
method described in \cite{venn01}.
We have used the coordinates of the \hi\ dynamical center,
position angle, and inclination obtained by \cite{bs98};
these values are also listed in Table~\ref{table_n6822}.
Some of the values of the deprojected galactocentric radii reported in
Table~\ref{table_alloxy} are different from those reported
in \cite{venn01} due to a transcription error in the latter work.
We note also that the \hii\ region identified as the ``nucleus''
in \cite{pes80} is likely to be the SNR Hodge~12.

We adopt $r_{\rm exp} = 3\farcm0 \pm 0\farcm1$ as the exponential scale
length from the carbon star survey by \cite{letarte02}.
Carbon stars are a good tracer of the intermediate- to old-age 
stellar populations, and we assume that the exponential scale
length derived by \cite{letarte02} is a fair measure of the
spatial extent of the underlying stellar population.
Thus, we have measured spectra of \hii\ regions out to radii
of almost 4.4 scale lengths.
\cite{mateo98} lists a scale length $2\farcm4 \pm 0\farcm4$, 
based on the observations by \cite{hodge91}.
Using the Hodge et al. value for the exponential scale length would
mean that we have covered \hii\ regions out to 5.5 scale lengths.

For a proper comparison with the present data, data for a number of
\hii\ regions were reanalyzed.
Abundances for Hodge~10, K$\gamma$, and Hubble~X were rederived using
the data from \cite{chandar00} and \cite{peimbert05};
our oxygen abundances agree with the published results.
Where there is more than one published measurement for an \hii\
region which we have also measured here, we adopt our derived oxygen
abundance, as indicated in Table~\ref{table_alloxy}.

\subsection{Is There An Abundance Gradient in NGC 6822?}

In Fig.~\ref{fig_gradient1}, we have plotted oxygen abundances
for all 19 nebulae in the present work as a function of physical
galactocentric radius. 
The intrinsic errors associated with bright-line abundances 
are of order 0.05~dex; in this plot, we have assigned 0.20~dex 
uncertainties from the bright-line method.
In addition to the stellar abundances from \cite{venn01},
we have also plotted oxygen abundances reported by \cite{smith75},
\cite{lequeux79}, \cite{pes80}, \cite{stm89}, \cite{rm95},
\cite{miller96}, \cite{chandar00}, \cite{hgom01}, \cite{lsv03},
and \cite{peimbert05}.
Oxygen abundances for \hii\ regions at radii $\la 0.5$~kpc
are in agreement with abundances derived for A-type supergiants
at similar radii.
Most of our derived oxygen abundances for \hii\ regions at
``intermediate'' radii ($\approx 1$~kpc; i.e., Hubble~V) agree
with previously reported values in the literature.
Our oxygen abundances for Hubble~I and Hubble~III also agree with the
values determined by \cite{pes80}.

There are, however, three data points which suggest significantly
lower oxygen abundances.
From \cite{chandar00}, they are K$\beta$, KD~12 (C9), and KD~20 (C10);
all three are labelled in Fig.~\ref{fig_gradient1}.
For KD~12, \cite{chandar00} reported weak \othree\ emission and
no \othreea.
Although we have not repeated a measurement for this nebula, simply
applying the bright-line method to their published line ratios in the
limit of zero reddening yields an oxygen abundance approximately
equal to 12$+$log(O/H) = 8.3, which is 0.5~dex higher than their
quoted value.
%
%
The larger oxygen abundance is in better agreement with abundances of
other nebulae at comparable radii.
Thus, we adopt a lower limit of 12$+$log(O/H) $\ga$ 7.8 for KD~12.
In K$\beta$, we measured \othreea, and the derived
oxygen abundance is consistent with the (bright-line) abundances
derived for \hii\ regions at similar radii (e.g., HK~16, HK~42,
KD~9).
We have subsequently adopted our \othreea\ oxygen abundance for
K$\beta$.
For KD~20, \othreea\ was neither detected in the present work
nor by \cite{chandar00}; thus, we have adopted a lower limit on the
oxygen abundance based on the [\ion{O}{2}] emission.

Oxygen abundances for the two planetary nebulae measured by
\cite{rm95} (Fig.~\ref{fig_n6822_ctr}; Table~\ref{table_alloxy}) are
slightly lower than the \hii\ region oxygen abundances at similar
radii.
However, planetary nebulae sample conditions that were present
about 1~Gyr ago in the interstellar medium, and the gas should be less
metal-rich at that time.
Accounting for the effects of mixing on the oxygen abundance
during the evolution of the progenitor star are additional
complications which are beyond the scope of the present work.
%
 
Various linear least-square fits to subsets of abundance data 
are described and shown in Table~\ref{table_fits} and
Fig.~\ref{fig_gradient2}.
When all of the 19 measured nebulae (out to four exponential scale
lengths) are included in the fit, no clear signature is found for a
radial abundance gradient. 
Including the three stellar abundances also produces zero slope.
A slope of $-0.14 \pm 0.07$ dex~kpc$^{-1}$ is obtained with a fit to
the five \hii\ regions with measured \othreea; the slope is consistent
with zero at $2\sigma$.
A fit to just the three A-type supergiants yields a 
significant slope; such a small sample can clearly bias
the resulting fit.
If we consider only the five \hii\ regions in the present data
with \othreea\ abundances and three additional \hii\ regions
with \othreea\ measurements from the literature (Hodge~10 and K$\gamma$
from \citealp{chandar00}; Hubble~X from \citealp{peimbert05}), we
obtain
\begin{equation}
12+ {\rm log(O/H)} = (8.23 \pm 0.05) \,+\, (-0.16 \pm 0.05) \, R,
\label{eqn_slope}
\end{equation}
where $R$ is the deprojected galactocentric radius in kpc, and the
slope is in units of dex~kpc$^{-1}$.
The fit corresponding to Equation~(\ref{eqn_slope}) is
seen in panel~(e) of Fig.~\ref{fig_gradient2} and is listed as
dataset~(e) in Table~\ref{table_fits}.
We remind the reader that despite only eight \hii\ regions,
we have culled the best available data at the present time.
This slope is slightly more significant ($3.2\sigma$), and
is about two times larger than that found for the Milky Way
\citep{rolleston00}, although the value approaches the range 
observed in other spiral galaxies (e.g., \citealp{vce92,zkh94}).
To convert the slope in units of dex~kpc$^{-1}$ to units of
dex~arcmin$^{-1}$ and dex~$r_{\rm exp}^{-1}$, we use the
linear-to-angular scale for a distance of 0.5~Mpc and the scale length
from Table~\ref{table_n6822}, and multiply the derived slopes in
Table~\ref{table_fits} by $0.144$ and $0.432$, respectively.
The slope above in Equation~(\ref{eqn_slope}) would thus be
$-0.023$ dex~arcmin$^{-1}$ and $-0.069$ dex~$r_{\rm exp}^{-1}$.
Finally, a fit to the eight \hii\ regions with \othreea\
oxygen abundances and three A-type supergiants gives a slope of $-0.24
\pm 0.07$ dex~kpc$^{-1}$ and a zero-point oxygen abundance equal
to 12$+$log(O/H) = $8.34 \pm 0.07$.
%

\subsection{Recent Studies}
\label{sec_recentwork}

\cite{vce92} included NGC~6822 in their sample of spiral galaxies to
examine the relationship between abundance gradients and other global
parameters.  
They used the data for seven \hii\ regions from \cite{smith75}
and \cite{pes80}, and determined a slope $-0.042$ dex~kpc$^{-1}$
and a central abundance of 12$+$log(O/H) = 8.24.
While their result is consistent with zero slope, these data 
were obtained with photon counters prone to nonlinear behavior
(i.e., \citealp{jenkins87}).

\cite{hgom01} suggested the existence of a radial gradient in nebular
oxygen abundances, based on small scale variations within the bright
\hii\ regions Hubble~V and Hubble~X.
While differences between \hii\ regions were not found, variations
as large as $\sim$ 0.4~dex were claimed on physical scales as small as
4~pc within a given \hii\ region.
These may be due to either true abundance variations or small-scale
temperature changes, but the authors could not distinguish
either scenario from their data.
Curiously, they report for Hubble~V a \halpha/\hbeta\ flux ratio of 
$1.50 \pm 0.07$, which is significantly lower than expected Balmer
ratios for typical conditions found in metal-poor \hii\ regions.
%

\cite{venn01} discussed the possibility of an abundance gradient,
based on their measurements of oxygen abundances from two 
A-type supergiants and \othreea\ measurements in \hii\ regions,
including the data from \cite{pes80} and \cite{stm89}.
They showed that the abundance gradient was  
$-0.18$ dex~kpc$^{-1}$ from a fit to the two A-type supergiants and 
the \othreea\ detections from \cite{pes80}.
%
%
\cite{pilyugin01b} reevaluated the metallicity-luminosity relation
for dwarf irregulars using his empirical method to derive oxygen
abundances.
He also examined the possibility of an abundance gradient in NGC~6822
by deriving empirical oxygen abundances using data from
\cite{lequeux79}, \cite{pes80}, \cite{stm89}, and \cite{hgom01}. 
Fitting the empirical abundances and the two stellar abundances
from \cite{venn01}, the resulting slope was $-0.035$ dex~kpc$^{-1}$,
claimed to be consistent with zero slope.

\cite{peimbert05} obtained high quality VLT spectra of the
\hii\ regions Hubble~V and Hubble~X.
They examined the data for possible temperature fluctuations
\citep{peimbert67}, which could give rise to small-scale chemical
inhomogeneities.
They observed recombination and collisionally-excited lines in
the giant \hii\ region Hubble~V.
They derived a recombination-line oxygen abundance
12$+$log(O/H) = $8.37 \pm 0.09$, which is in better agreement 
with the mean oxygen abundance for the two A-type supergiants
(12$+$log(O/H) = $8.36 \pm 0.19$; \citealp{venn01}).
From the collisionally-excited emission lines, they derived
12$+$log(O/H) = $8.08 \pm 0.03$ and $8.34 \pm 0.06$ for zero and
non-zero temperature fluctuations, respectively.
Our direct oxygen abundance ($8.14 \pm 0.05$) agrees with their
collisional-line abundance derived with zero temperature 
fluctuations.
For Hubble~X, they were unable to observe the oxygen recombination
lines, and derived oxygen abundances from collisionally-excited lines
of $8.01 \pm 0.05$ and $8.19 \pm 0.16$ for zero and non-zero
temperature fluctuations, respectively.
We rederived oxygen abundances using their reported fluxes 
with our two-zone model and with zero temperature fluctuations.
We obtained 12$+$log(O/H) = 8.11 and 8.06 
for Hubble~V and Hubble~X, respectively, which agree
with their published values. 
However, for Hubble~V we derived $T_e$(O$^+$)
= $11200 \pm 750$~K, while they reported $13000 \pm 1000$~K.
This is a significant difference, especially considering
that the values of $T_e$(O$^{+2}$) are in excellent agreement.
It is interesting to note that their N/O values 
(log(N/O) $\approx -1.3$) are more in line with the values expected
for their oxygen abundances in comparison to other metal-poor
galaxies.
To compare the present nebular data with other published work, we have
only considered their collisional-line abundances, and assumed zero
temperature fluctuations.
We note that \cite{carigi05} have recently described a suite of
chemical evolution models, which agree with the photometric properties
and the star-formation history of NGC~6822. 

\subsection{Abundance Variations: The Dispersion in Oxygen Abundance}
\label{sec_variations}



If we assume zero radial gradient, we derive a mean nebular oxygen
abundance for the \hii\ regions with \othreea\ measurements.
We obtained \othreea\ detections in both grating settings
for \hii\ regions Hubble~V, K$\alpha$, and KD28e; we computed
a single average for each of these three \hii\ regions.
The mean oxygen abundance for Hubble~V, K$\alpha$,
K$\beta$, Hubble~I, and K$\beta$ is
(O/H) = $(1.29 \pm 0.30) \times 10^{-4}$, or 
12$+$log(O/H) = $8.11 \pm 0.10 (^{+0.08}_{-0.14})$. 
For the logarithmic value, the two errors represent the overall
dispersion in the mean, and the distribution of minimum and maximum
values, respectively.
The mean value corresponds to [O/H] = $-0.55$~dex,
or $28$\% of the solar value.
For historical completeness, our adopted mean nebular oxygen abundance
would correspond to [O/H] = $-0.83$ for a solar value of 
12$+$log(O/H) = 8.93 stated by \cite{ag89}.
If we include the three additional \hii\ regions with recent
\othreea\ measurements (Hodge~10 and K$\gamma$ from
\citealp{chandar00}; Hubble~X from \citealp{peimbert05}),
the mean oxygen abundance is 12$+$log(O/H) = 
$8.08 \pm 0.09 (^{+0.11}_{-0.10})$.
The mean oxygen abundance is consistent with abundances for
nearby dwarf irregular galaxies at comparable optical luminosities
(e.g., \citealp{skh89,rm95,lee03field}).
A reexamination of the various fits in Fig.~\ref{fig_gradient2} shows
that the largest dispersion ($\la$ 0.2~dex) occurs when the
bright-line abundances are included in the fit which results
in zero slope.
This is no surprise, as the relative differences between
direct and bright-line abundances is about 0.2~dex 
(see Fig.~\ref{fig_oxydiff}).
This argues for relative chemical homogeneity to the level
of about 0.2~dex over scales spanning over four exponential scale
lengths ($\ga$ 2~kpc).

However, we have seen that a subset with the highest quality
(\othreea\ measurements) suggests an abundance gradient.
Is the gradient related to the extended spatial
distributions of \hi\ gas and blue stars?
The blue stars are too old ($\approx$ 100--200 Myr; 
\citealp{debw03,komiyama03}) to affect the most recent chemical
enrichment found in \hii\ regions.
Is the gradient caused by \hi\ dilution at large galactocentric
radius?
The \hi\ cloud to the northwest of NGC~6822 and the overall
unusual spatial distribution of \hi\ \citep{debw00,debw05}
may be diluting the present-day chemical enrichment in the outlying
regions of the galaxy, which could give rise to the gradient
observed.

Additional measurements would clearly be valuable, although there
are no \hii\ regions as bright as Hubble~V at large galactocentric
radii. 
While the \hii\ region KD~28 ($R/r_{\rm exp} \approx 4.4$) anchors
the slope at large galactocentric radii, only bright nebular emission
lines were observed for this faint diffuse \hii\ region. 
What is very interesting is that KD~28 is located where a spur of
young blue stars extends from the southeast corner of
the optical bar (e.g., \citealp{debw03,komiyama03}).
A closer inspection of the Local Group Survey \halpha\ image 
reveals a number of {\em compact\/} emission-line regions in the
vicinity of KD~28.
\cite{debw05} have found additional \hii\ regions at intermediate
radii from very deep \halpha\ imaging with the INT 2.5-m telescope.
Very deep spectra would be required to detect \othreea\ for KD~28 (or
any of the other \hii\ regions). 
These data could help secure additional evidence of an abundance
gradient at large galactocentric radii.

Spectroscopy of other stellar probes would also provide valuable
information. 
Data for two additional A-type supergiants at intermediate
galactocentric radii, which would better bridge the measurements of
the other stars at small radii, will be presented in a subsequent
paper (Venn et al., in preparation).
Additional candidates of A-type supergiants for subsequent
spectroscopy would clearly be valuable as an additional probe of the
most recent episode of chemical enrichment.
Spectroscopy of the confirmed PNe candidates (e.g., \citealp{leisy05})
would also indicate the consistency of abundances at a given radius,
and would show whether the interstellar medium about 1~Gyr ago also
exhibited overall spatial chemical homogeneity (modulo possible mixing
in the progenitor stars). 
If they are sufficiently luminous, spectroscopy of individual blue
stars at large galactocentric radii could provide another valuable
test, as the blue stars have ages intermediate to the young
\hii\ regions and the (relatively) older PNe.
\cite{tolstoy01} obtained direct spectroscopic metallicities
for 23 red giant stars in NGC~6822 and determined the metallicity
distribution function with mean [Fe/H] = $-1 \pm 0.5$.
This mean iron abundance is similar to the photometric iron abundances
of stars \citep{lee93,gallart96b,gallart96c} and to the spectroscopic
abundances of open clusters \citep{cb98} in the galaxy.
We have plotted spectroscopic iron abundances of the 23 red giant 
stars as a function of their galactocentric radius in
Fig.~\ref{fig_iron}. 
There is no signature of a radial gradient in iron abundance for the 
older stellar population.
This is not entirely surprising, as the red giant stars have a larger
range of (old) ages than the \hii\ regions.
The larger range in age ``hides'' previous episodes of star
formation and the subsequent episodes of chemical enrichment.
Continuing work with the VLT for an additional 90 red giant stars in
NGC~6822 (A.~A. Cole et al., in preparation) could provide stronger
constraints on the history of star formation over a larger range of
ages.

\section{Conclusions}		
\label{sec_concl}		

Optical spectra were measured for 19 nebulae in NGC~6822 
at galactocentric radii between 0.05 kpc and 2 kpc, or out
to about four exponential scale lengths.
\othreea\ was detected in five \hii\ regions, and subsequent
direct oxygen abundances were derived.
Oxygen abundances for the remaining \hii\ regions were derived using
bright-line methods.
Oxygen abundances for the A-type supergiant stars are consistent
with \hii\ region abundances at comparable galactocentric radii.
Linear least-square fits to various subsets of abundance data 
were obtained.
When all of the measured nebulae are included, no clear signature is
found for an abundance gradient out to over four exponential scale
lengths.
A fit to \hii\ regions with only \othreea\ detections is consistent
with zero slope ($2\sigma$).
The abundance gradient becomes slightly more significant 
($3.2\sigma$) when three additional \hii\ regions with \othreea\
measurements from the literature are included.
The resulting slope and extrapolated central abundance are
$-0.16 \pm 0.05$ dex~kpc$^{-1}$ and 12$+$log(O/H) = $8.23 \pm 0.05$,
respectively.
Assuming zero abundance gradient, we take the five \hii\
regions with \othreea\ detections from our present work, and derive a
mean nebular oxygen abundance 12$+$log(O/H) = $8.11 \pm 0.10$, which
corresponds to [O/H] = $-0.55$.
Additional deep high-quality spectra of nebulae and stars are required
to distinguish clearly between either a zero or a non-zero slope.
The latter would be confirmation of an abundance gradient seen
for the first time in a dwarf irregular galaxy.
%

\begin{acknowledgements}	

We thank the anonymous referee for their careful reading and
their helpful comments which improved the presentation of the
manuscript. 
H.~L. thanks 
Fabio Bresolin, John Cannon, Maria-Rosa Cioni, Andrew Cole, 
Erwin de Blok, Ana Maria Hidalgo-G\'amez, 
Pierre Leisy, Laura Magrini, Lissa Miller, Erik Muller, Michael Richer,
Jeroen Stil, and Marc Zimer for helpful and informative exchanges.
We thank Antonio Peimbert, Manuel Peimbert, and Leticia Carigi for 
copies of their manuscripts before publication.
We are grateful to ESO for awarded telescope time, and H.~L. thanks
Lisa Germany and the staff at ESO La Silla for their help in acquiring
the data.
H.~L. and E.~D.~S. acknowledge partial support from a NASA LTSARP grant
NAG~5--9221 and the University of Minnesota.
K.~A.~V. thanks the National Science Foundation for support through
a CAREER award AST~99--84073. 
For their one-year visit, E.~D.~S. and K.~A.~V. thank the Institute 
of Astronomy, University of Cambridge for their hospitality and support.
Some data were accessed as Guest User, Canadian Astronomy Data Center,
which is operated by the Dominion Astrophysical Observatory for the 
National Research Council of Canada's Herzberg Institute of
Astrophysics.
This research has made use of NASA's Astrophysics Data System, and
of the NASA/IPAC Extragalactic Database, which is operated by
the Jet Propulsion Laboratory, California Institute of Technology,
under contract with the National Aeronautics and Space Administration. 

\vspace*{2mm}
{\it Facility:} ESO:3.6m (EFOSC2)

\end{acknowledgements}

\clearpage	

\begin{figure}
\figurenum{1}
\caption{
%
NGC 6822 : the field of view is about 21\farcm9 $\times$ 19\farcm2.
North is at the top and East is to the left.
White objects on the image indicate bright sources
in the unsubtracted \halpha\ image from the Local Group Survey
\citep{massey_lgs}.
Numeric labels indicate subsequent fields of view around Hubble~I and
III, Hubble~V, and the center in Figs.~\ref{fig_n6822_hu3} to
\ref{fig_n6822_ctr}, respectively.
Three A-type supergiant stars and two planetary nebulae in the
galaxy are indicated with blue and yellow symbols, respectively.
The \hi\ dynamical center of NGC~6822 is marked with a cross
\citep{bs98}, and the 10\arcmin\ by 7\arcmin\ (1.4 kpc by 1.0 kpc)
ellipse at a position angle of 112\degr\ indicates approximately the
orientation of the \hi\ distribution, which extends much farther
east-west than what can be shown here (e.g., \citealp{debw00}).
}
\label{fig_n6822_gxy}
\end{figure}

\begin{figure}
\figurenum{2}
\caption{
Long-slit position for blue spectra of \hii\ regions Hubble~III
(Hu~III) and Hubble~I (Hu~I).
The field of view is 2\farcm7 $\times$ 2\farcm4.
Orientation is the same as in Figure~\ref{fig_n6822_gxy}.
Resolved features within Hu~III and Hu~I are labeled for which
spectra were obtained. 
The separation between the solid lines corresponds approximately
to the 1\farcs5 slit width projected on the sky.
The location of an A-type supergiant star, CW~22, in NGC~6822
is indicated.
}
\label{fig_n6822_hu3}
\end{figure}

\begin{figure}
\figurenum{3}
\caption{
Long-slit positions for spectra in the area around
the giant \hii\ region Hubble~V (Hu~V).
The field of view is 2\farcm7 $\times$ 2\farcm4 and
the orientation is the same as in Figure~\ref{fig_n6822_gxy}.
The long-slit orientations on 26~Aug (grating 11) and
on 31~Aug (grating 7) are labeled.
For a given slit position, the separation between the solid lines 
corresponds approximately to the 1\farcs5 slit width projected 
on the sky.
Labeled are \hii\ regions for which spectra were obtained
and presented in this work, except Hodge~6 and KD~12 (C9 in
\citealp{chandar00}).
The locations of the latter two \hii\ regions are provided for 
clarity.
}
\label{fig_n6822_hu5}
\end{figure}

\begin{figure}
\figurenum{4}
\caption{
%
The central 5\farcm4 $\times$ 4\farcm8 field of view of NGC~6822.
Orientation is the same as in Figure~\ref{fig_n6822_gxy}.
\hii\ regions, stars, and planetary nebulae are labelled; see
Table~\ref{table_alloxy}.
KD~20 or C10 as labeled by \cite{chandar00} is indicated.
Two A-type supergiant stars, CW173 and CW175,
\citep{venn01}, and two planetary nebulae, S16 and S33 \citep{rm95}
are marked.
The \hi\ dynamical center \citep{bs98} is marked with a cross.
}
\label{fig_n6822_ctr}
\end{figure}

\begin{figure}
\figurenum{5}
\caption{
Blue spectra of NGC~6822 nebulae: \othreea\ detections.
The observed flux per unit wavelength is plotted versus wavelength.
Within each panel, the full spectrum and an expanded view of the
spectrum to highlight faint emission lines are shown.
The \othreea\ line is indicated by an arrow in each panel.
}
\label{fig_gr07spc}
\end{figure}

\begin{figure}
\figurenum{6}
\caption{
Low-dispersion spectra of NGC~6822 nebulae: 
\othreea\ detections are shown in the first three panels.
Comments are the same as in Fig.~\ref{fig_gr07spc}.
Note that \hii\ regions KD~28e and KD~28 are not located
in the same part of the galaxy.
KD~28 is located in the southern periphery of the galaxy
(Fig.~\ref{fig_n6822_gxy}), whereas 
KD~28e is located near the center of the galaxy
(Fig.~\ref{fig_n6822_ctr}).
The spectrum for KD~28 is shown, because this \hii\ region is 
farthest from the center with a measurement, and anchors the
fit at large radii (see Figs.~\ref{fig_gradient1} and
\ref{fig_gradient2}).
}
\label{fig_gr11spc}
\end{figure}

\begin{figure}
\figurenum{7}
\caption{
Two-dimensional spectrum image from \otwo\ to \otwored\ : 
spectra for KD~24, HK~69, and Hodge~12 (Ho~12) are shown.
Dark objects on the image indicate bright sources.
Key emission lines are labelled; vertical stripes are
night-sky lines.
The long-slit was positioned so that north was at the top
of the frame.
Resolved structure is evident in the emission lines
for Ho~12.
}
\label{fig_snr2d}
\end{figure}

\begin{figure}
\figurenum{8}
\epsscale{0.7}
\plotone{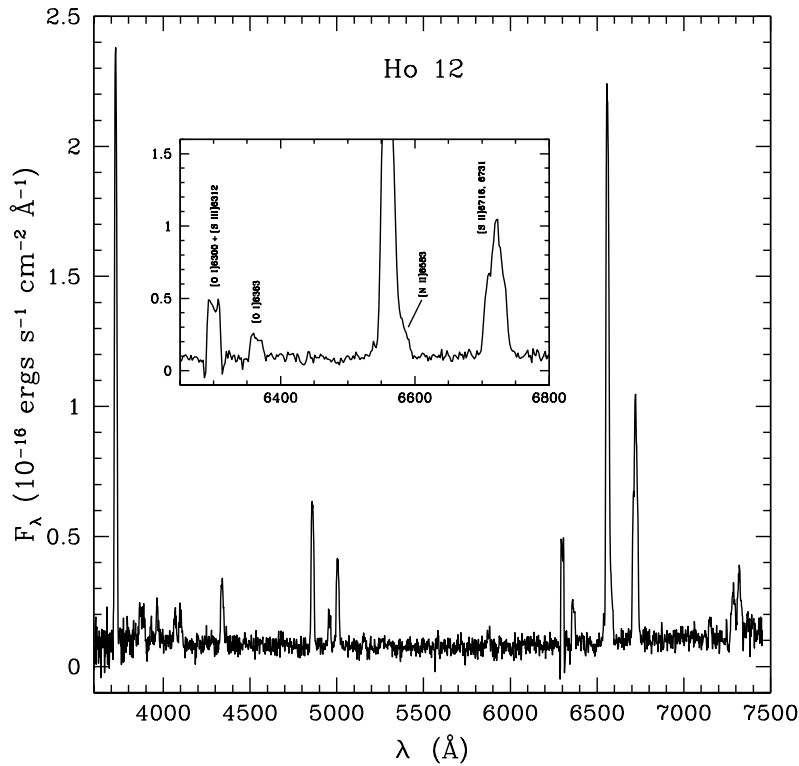}
\caption{
Low-dispersion spectrum of the center of Hodge~12.
The inset shows the spectrum in the wavelength range between
6250 and 6800~\AA, highlighting relatively bright
\ooneb\ and \stwo\ emission lines.
}
\label{fig_snr}
\end{figure}

\begin{figure}
\figurenum{9}
\caption{
Monte Carlo simulations of solutions for the reddening, $c(\hbeta)$,
and the underlying Balmer absorption with equivalent width, 
EW$_{\rm abs}$, from hydrogen Balmer flux ratios.
Dotted lines mark zero values for each quantity.
The results here are shown for the low-dispersion
spectrum of Hubble~V.
Each small point is a solution derived from a different realization
of the same input spectrum.
The large filled circle with error bars shows the mean result with
$1\sigma$ errors derived from the dispersion in the solutions.
In this example, the Monte Carlo simulations have yielded an
unphysical result for the underlying Balmer absorption; thus, we adopt
an underlying Balmer absorption with equivalent width equal to zero.
}
\label{fig_monte}
\end{figure}

\begin{figure}
\figurenum{10}
\epsscale{0.63}
\plotone{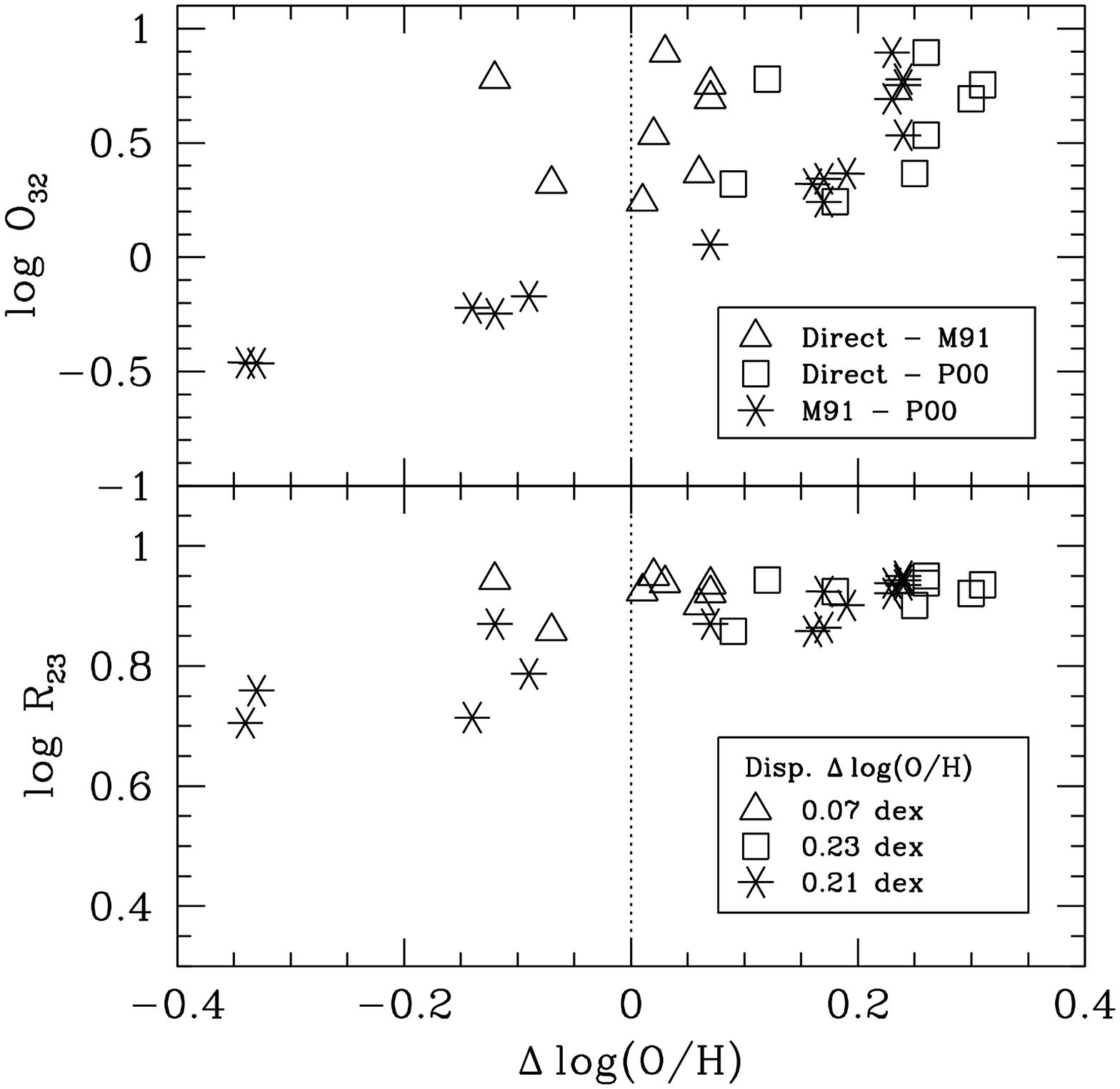}
\caption{
Difference in oxygen abundance from various methods
versus log~$O_{32}$ (top panel), and 
versus log~$R_{23}$ (bottom panel).
Each symbol represents an \hii\ region.
``Direct'' denotes oxygen abundances derived from \othreea\
measurements, ``M91'' denotes oxygen abundances derived using
the bright-line method by \cite{mcgaugh91}, and ``P00'' denotes
oxygen abundances derived using the bright-line method 
by \cite{pilyugin00}.
Vertical dotted lines in both panels mark zero differences in 
oxygen abundance.
Dispersions in abundance differences are indicated in the legend
of the bottom panel.
In the absence of \othreea, oxygen abundances derived with
the bright-line method are accurate to within $\approx$ 0.2~dex.
}
\label{fig_oxydiff}
\end{figure}

\begin{figure}
\figurenum{11}
\epsscale{0.6}
\plotone{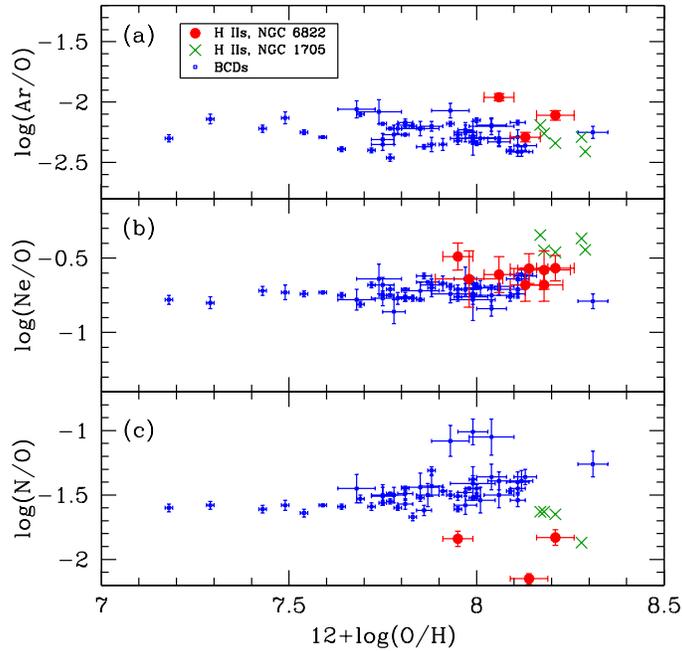}
\caption{
Nebular abundance ratios versus oxygen abundance: 
(a) argon-to-oxygen, (b) neon-to-oxygen, and (c) nitrogen-to-oxygen.
Filled circles indicate \hii\ regions in NGC~6822 with 
\othreea\ measurements.
Small squares represent blue compact dwarf galaxies \citep{it99}.
Crosses indicate \hii\ regions in the starbursting
dwarf galaxy NGC~1705 \citep{ls04}.
}
\label{fig_zratios}
\end{figure}

\begin{figure}
\figurenum{12}
\epsscale{0.67}
\plotone{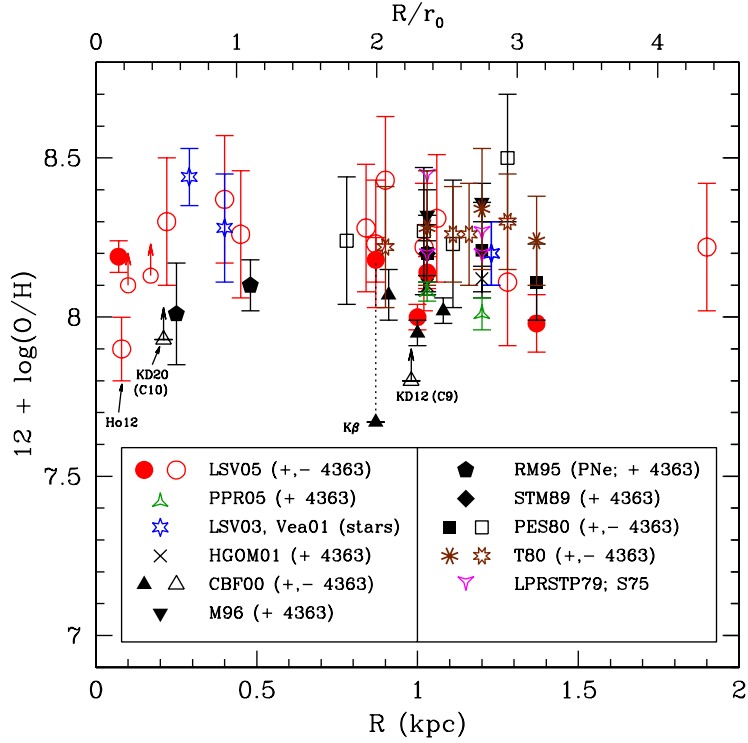}
\caption{\footnotesize 
Oxygen abundance versus deprojected galactocentric radius.
The bottom horizontal axis shows the physical radius in kpc, whereas
the top horizontal axis shows the radius normalized by the exponential 
scale length, $r_0$, reported by \cite{letarte02}.
Symbols represent data from the following sources:
LSV05 -- nebulae from this work;
PPR05 -- \hii\ regions from \cite{peimbert05};
LSV03, Vea01 -- A-type supergiants from \cite{venn01}, \cite{vm02},
and \cite{lsv03};
HGOM01 -- \hii\ regions from \cite{hgom01};
CBF00 -- \hii\ regions from \cite{chandar00};
M96 -- \hii\ regions from \cite{miller96};
RM95 -- planetary nebulae from \cite{rm95};
STM89 -- \hii\ region from \cite{stm89};
PES80 -- \hii\ regions from \cite{pes80};
T80 -- \hii\ regions from \cite{talent80}; 
LPRSTP79 -- \hii\ regions from \cite{lequeux79}; and
S75 -- \hii\ regions from \cite{smith75}.
The legend indicates the various symbols representing
positive \othreea\ detections in \hii\ regions.
%
%
The three anomalously low oxygen abundances reported by
\cite{chandar00} for K$\beta$, KD12 (C9), and KD20 (C10)
are discussed in the text.
A dotted vertical line connects two measurements of the
\hii\ region K$\beta$; see text.
Also labeled is the supernova remnant Hodge 12 (Ho12).
}
\label{fig_gradient1}
\end{figure}

\begin{figure}
\figurenum{13}
\epsscale{0.7}
\plotone{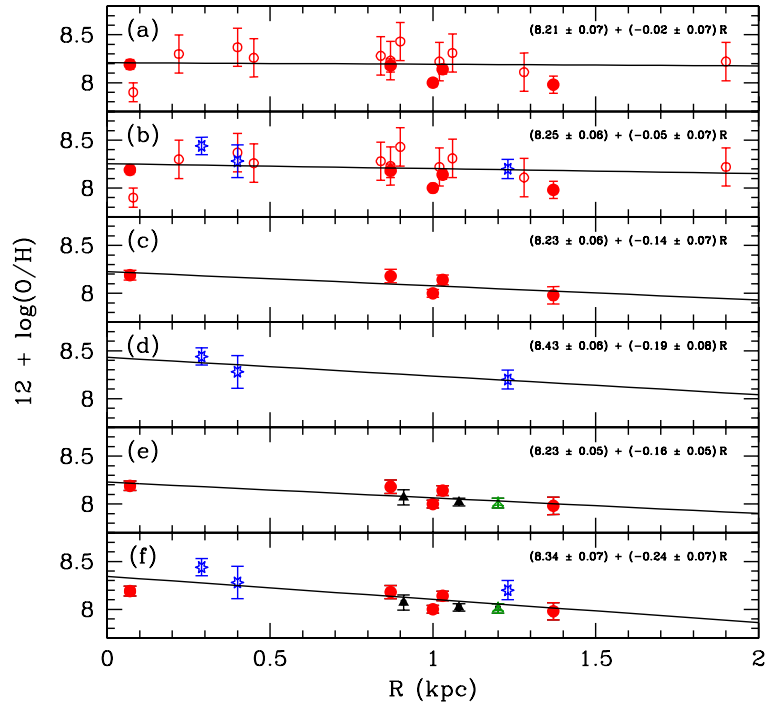}
\caption{
Fits to subsets of oxygen abundance data.
Symbols and the plot axes for each panel are the same as in
Fig.~\ref{fig_gradient1}.
Each panel corresponds to the specific subset of abundance data, and
the linear least-squares fit described in Table~\ref{table_fits}.
For nebular \othreea\ abundances, we only considered data
obtained with CCDs for an appropriate comparison.
We adopt oxygen abundances derived for all \hii\ regions
which were remeasured in the present work (e.g., our O/H for Hubble~V).
}
\label{fig_gradient2}
\end{figure}

\begin{figure}
\figurenum{14}
\epsscale{0.7}
\plotone{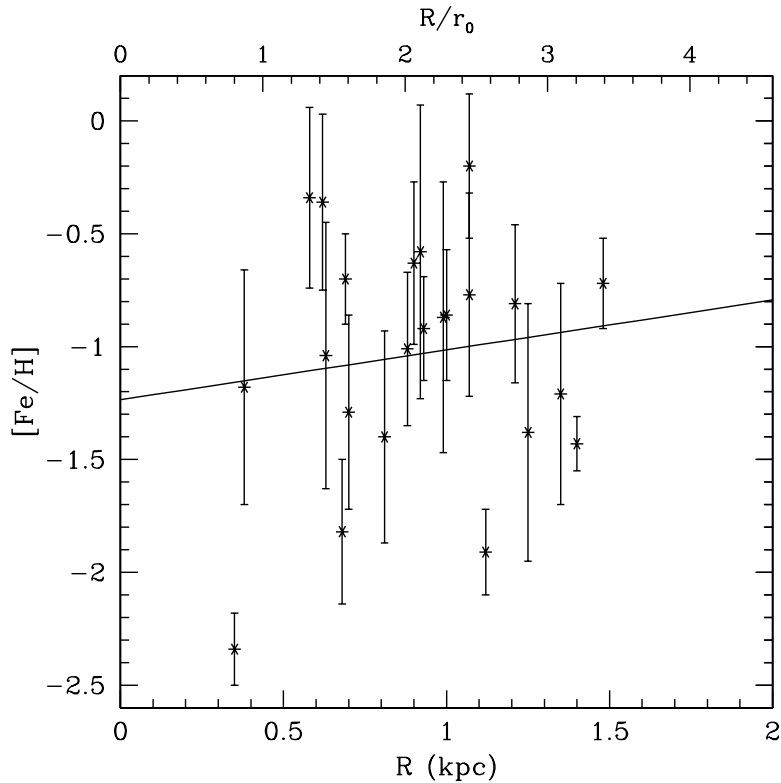}
\caption{
Iron abundance, [Fe/H], versus deprojected galactocentric radius, $R$,
for 23 red giant stars in the northern half of NGC~6822
\citep{tolstoy01}. 
The resulting least-squares fit is expressed as
[Fe/H] = $(-1.23 \pm 0.34) + (0.22 \pm 0.35) R$.
}
\label{fig_iron}
\end{figure}

\clearpage	

\begin{table}
\setlength{\tabcolsep}{2mm}
\tablenum{1}
\begin{center}
\renewcommand{\arraystretch}{1.}
\caption{
Basic data for NGC~6822.
\label{table_n6822}
}
\begin{tabular}{ccc}
\tableline \tableline
Property & Value & References \\
\tableline
Morphological Type & IB(s)m & \nodata \\
Alternate Names & DDO 209, IC 4895 & \nodata \\
Right ascension (J2000)\tablenotemark{a} & 
	\phs$19^h\;44^m\;56\fs4$& 1 \\
Declination (J2000)\tablenotemark{a} & 
	$-14\arcdeg\;48\arcmin\;04\farcs5$ & 1 \\
Position angle & 112\arcdeg & 1 \\
Inclination & 50\fdg1 & 1 \\
Distance & 0.47--0.50 Mpc & 2, 3, 4 \\
Linear to angular scale at this distance & 
	2.3--2.4 pc arcsec$^{-1}$ & 5 \\
Heliocentric velocity & $-54$ km s$^{-1}$ & 6 \\
$r_{\rm exp}$, exponential scale length & 
	$3\farcm0 \pm 0\farcm1$ & 7 \\
$D_{25}\,$\tablenotemark{b} & 15\farcm5 & 8 \\
$B_T\;$\tablenotemark{c} & 9.32 & 8 \\
$E(B-V)\;$\tablenotemark{c} & 0.236 & 8 \\
$M_{B,0}\;$\tablenotemark{c} & $-15.1$ & 5 \\ 
$F_{21}$, 21-cm flux integral & 2399 Jy km s$^{-1}$ & 8, 9 \\
$<12+$log(O/H)$>$, A-sg$\,$\tablenotemark{d} & 
	$8.36 \pm 0.19$ ($\pm$ 0.21) & 10 \\
$<12+$log(O/H)$>$, \hii$\,$\tablenotemark{e} & 
	$8.11 \pm 0.10$ & 5 \\
$\Delta[$log(O/H)$]/\Delta R\,$\tablenotemark{f} & 
	$-0.16 \pm 0.05$ dex kpc$^{-1}$ & 5 \\
12$+$log(O/H)$_c\,$\tablenotemark{f} & 
	$8.23 \pm 0.05$ & 5 \\
\tableline
\end{tabular}
\vspace*{2mm}
\tablenotetext{a}{
Coordinates of the \hi\ dynamical center.
}
\tablenotetext{b}{
Major angular diameter measured at 25 mag~arcsec$^{-2}$.
}
\tablenotetext{c}{
Apparent total $B$ magnitude,
foreground reddening to the galaxy,
and reddening-corrected absolute total $B$ magnitude
(0.5~Mpc distance), respectively.
}
\tablenotetext{d}{
Mean stellar oxygen abundance from two A-type supergiant stars.
}
\tablenotetext{e}{
Zero gradient: mean nebular oxygen abundance derived from \othreea\
measurements in five \hii\ regions.
}
\tablenotetext{f}{
Non-zero gradient ($3.2\sigma$): slope and extrapolated central oxygen
abundance, respectively; see dataset e from Table~\ref{table_fits}.
}
\tablerefs{\footnotesize
(1) \cite{bs98}; 
(2) \cite{gallart96a}; 
(3) \cite{clementini03}; 
(4) \cite{pietr04}; 
(5) present work;
(6) \cite{mateo98}; 
(7) \cite{letarte02}; 
(8) \cite{kara04locvol}; 
(9) \cite{hr86}; 
(10) \cite{venn01}. 
}
\end{center}
\end{table}

\begin{table}
\tablenum{2}
\begin{center}
\renewcommand{\arraystretch}{1.1}
\caption{
Properties of EFOSC2 spectrograph employed at the ESO La Silla
3.6-m telescope.
\vspace*{3mm}
\label{table_obsprops}
}
\begin{tabular}{ccc}
\tableline \tableline
\multicolumn{3}{c}{{\sf Loral CCD (No. 40)}} \\ 
\tableline
Total area & \multicolumn{2}{c}{2048 $\times$ 2048 pix$^2$} \\
Field of view & \multicolumn{2}{c}{5\farcm2 $\times$ 5\farcm2} \\
Pixel size & \multicolumn{2}{c}{15 $\mu$m} \\
Image scale & \multicolumn{2}{c}{0\farcs16 pixel$^{-1}$} \\
Gain & \multicolumn{2}{c}{1.3 $e^-$ ADU$^{-1}$} \\ 
Read-noise (rms) & \multicolumn{2}{c}{9 $e^-$} \\ 
\tableline
\multicolumn{3}{c}{{\sf Long slit}} \\ \tableline
Length & \multicolumn{2}{c}{$\simeq 5$\arcmin} \\
Width & \multicolumn{2}{c}{1\farcs5} \\
\tableline
& {\sf Grating 7} & {\sf Grating 11} \\ 
& {\sf (``blue'')} & {\sf (``low dispersion'')} \\ \tableline
Groove density & 600 lines mm$^{-1}$ & 300 lines mm$^{-1}$ \\
Blaze $\lambda$ (1st order) & 3800 \AA & 4000 \AA \\
Dispersion & 0.96 \AA\ pixel$^{-1}$ & 2.04 \AA\ pixel$^{-1}$ \\
Effective $\lambda$ range & 3270--5240 \AA & 3380--7520 \AA  \\ 
\tableline
\end{tabular}
\end{center}
\end{table}

\begin{table}
\tablenum{3}
\setlength{\tabcolsep}{2mm}
\begin{center}
\renewcommand{\arraystretch}{1.1}
\caption{
Log of Observations.
\vspace*{3mm}
\label{table_obslog}
}
\begin{tabular}{cccccccccc}
\tableline \tableline
& Alternate & Date & & & $t_{\rm total}$ & & $|\Delta\theta|$ & & RMS \\ 
\hii\ Region & Name(s) & (UT 2003) & Spectra & $N_{\rm exp}$ & (s) & 
$\langle X \rangle$ & ($^{\circ}$) & \othreea & (mag) \\
(1) & (2) & (3) & (4) & (5) & (6) & (7) & (8) & (9) & (10) \\
\tableline
HK 16 & \nodata & 26 Aug & Low disp. & 3 $\times$ 1200 & 3600 & 1.11 &
8 & \nodata & 0.030 \\ 
HK 42 & \nodata & 26 Aug & Low disp. & 3 $\times$ 1200 & 3600 & 1.11 &
8 & \nodata & 0.030 \\ 
HK 69 & \nodata & 28 Aug & Low disp. & 3 $\times$ 1200 & 3600 & 1.03 &
20 & \nodata & 0.034 \\ 
HK 70 & \nodata & 28 Aug & Low disp. & 3 $\times$ 1200 & 3600 & 1.03 &
83 & \nodata & 0.034 \\
Hodge 7 & KD 11 & 27 Aug & Low disp. & 6 $\times$ 1200 & 7200 & 1.11 &
20 & \nodata & 0.029 \\ 
Hodge 12 & KD 23 & 28 Aug & Low disp. & 3 $\times$ 1200 & 3600 & 1.03 &
20 & \nodata & 0.034 \\
Hodge 12 & KD 23 & 31 Aug & Blue & 2 $\times$ 1200 & 2400 & 1.04 &
13 & \nodata & 0.025 \\
Hubble I & Hodge 2; KD 1 & 31 Aug & Blue  & 1 $\times$ 1200 & 1200 & 
1.09 & 82 & yes & 0.025 \\ 
Hubble III & Hodge 4; KD 4 & 31 Aug & Blue & 1 $\times$ 1200 & 1200 &
1.09 & 82 & \nodata & 0.025 \\ 
Hubble V & Hodge 9, 11; KD 19 & 26 Aug & Low disp. & 3 $\times$ 1200 &
3600 & 1.11 & 8 & yes & 0.030 \\ 
Hubble V & Hodge 9, 11; KD 19 & 31 Aug & Blue & 1 $\times$ 1200 & 1200 &
1.14 & 68 & yes & 0.025 \\ 
K$\alpha$ & KD 2e & 28 Aug & Low disp. & 3 $\times$ 1200 & 3600 & 1.12 &
14 & yes & 0.034 \\ 
K$\alpha$ & KD 2e & 31 Aug & Blue & 2 $\times$ 1200 & 2400 & 1.04 & 
42 & yes & 0.025 \\ 
K$\beta$ & KD 5e & 28 Aug$\,$\tablenotemark{a} & Low disp. & 
3 $\times$ 1200 & 3600 & 1.12 & 14 & \nodata & 0.034 \\
K$\beta$ & KD 5e & 31 Aug$\,$\tablenotemark{b} & Blue & 2 $\times$ 1200 &
2400 & 1.04 & 42 & yes & 0.025 \\
KD 9 & \nodata & 26 Aug & Low disp. & 3 $\times$ 1200 & 3600 & 1.11 &
8 & \nodata & 0.030 \\ 
KD 20 & C10$\,$\tablenotemark{c} & 28 Aug & Low disp. &
3 $\times$ 1200 & 3600 & 1.09 & 83 & \nodata & 0.034 \\
KD 21 & \nodata & 26 Aug & Low disp. & 3 $\times$ 1200 & 3600 & 1.11 &
8 & \nodata & 0.030 \\
KD 22e & \nodata & 31 Aug & Blue & 1 $\times$ 1200 & 1200 & 1.14 &
68 & \nodata & 0.025 \\ 
KD 24 & \nodata & 28 Aug & Low disp. & 3 $\times$ 1200 & 3600 & 1.03 &
20 & \nodata & 0.034 \\ 
KD 25 & \nodata & 26 Aug & Low disp. & 6 $\times$ 1200 & 7200 & 1.04 &
44 & \nodata & 0.030 \\
KD 28 & \nodata & 27 Aug & Low disp. & 6 $\times$ 1200 & 7200 & 1.07 &
54 & \nodata & 0.029 \\
KD 28e & \nodata & 28 Aug & Low disp. & 3 $\times$ 1200 & 3600 & 1.09 & 
83 & yes & 0.034 \\
KD 28e & \nodata & 31 Aug & Blue & 2 $\times$ 1200 & 2400 & 1.04 & 
13 & yes & 0.025 \\
\tableline
\end{tabular}
\vspace*{2mm}
\tablenotetext{a}{
The long-slit was placed through the center of the diffuse emission
in K$\beta$.
}
\tablenotetext{b}{
The long-slit was placed through the bright feature to the northwest
in K$\beta$.
}
\tablenotetext{c}{
The coordinates for KD 20 are the same for the object identified 
as ``C10'' by \cite{chandar00}.
}
\tablecomments{
Cols.~(1) and (2): \hii\ region, arranged in alphabetical order by
their primary name and alternate names, respectively; see \cite{hkl88}
for description.
KD numbers followed by an ``e'' were identified as ``stellar'' objects
in Table~3 of \cite{kd82}.
Col.~(3): Date of observation.
Col.~(4): Spectra type; see Table~\ref{table_obsprops}.
Col.~(5): Number of exposures obtained and the length of each
exposure in seconds. 
Col.~(6): Total exposure time.
Col.~(7): Mean effective airmass.
Col.~(8): Average departure of the observing angle from the
parallactic angle.
Col.~(9): \othreea\ detection.
Col.~(10): Relative root--mean--square error in the sensitivity
function obtained from observations of standard stars.
}
\end{center}
\end{table}

\begin{table}
\tablenum{4a}
\setlength{\tabcolsep}{2mm}
\begin{center}
\renewcommand{\arraystretch}{1.0} 
\caption{
Blue spectra: line ratios and properties for nebulae
Hodge~12, Hubble~I, and Hubble~III.
\vspace*{3mm}
\label{table_gr07data1}
}
\begin{tabular}{rccccccc}
\tableline \tableline
& & & &
\multicolumn{2}{c}{Hu I} & 
\multicolumn{2}{c}{Hu I} \\ 
& & \multicolumn{2}{c}{Ho 12} & 
\multicolumn{2}{c}{E ctr} & 
\multicolumn{2}{c}{NW spot} \\ 
\multicolumn{1}{c}{Property} &
\multicolumn{1}{c}{$f(\lambda)$} &
\multicolumn{1}{c}{$F$} & \multicolumn{1}{c}{$I$} &
\multicolumn{1}{c}{$F$} & \multicolumn{1}{c}{$I$} &
\multicolumn{1}{c}{$F$} & \multicolumn{1}{c}{$I$} \\
\tableline
$[\rm{O\;II}]\;3727$ & $+0.325$ &
	$341.2 \pm 4.7$ & $444.2 \pm 6.1$ &
	$222.6 \pm 5.0$ & $228.0 \pm 6.2$ &
	$235.4 \pm 5.1$ & $233.5 \pm 5.1$
\\
$[{\rm Ne\;III}]\;3869$ & $+0.294$ &
	$20.3 \pm 1.9$ & $25.8 \pm 2.4$ &
	$26.9 \pm 3.9$ & $27.5 \pm 4.0$ &
	$25.6 \pm 3.9$ & $25.4 \pm 3.9$
\\
${\rm H}8 + {\rm He\;I}\;3889$ & $+0.289$ &
	$22.5 \pm 2.2$ & $28.8 \pm 2.8$ &
	$19.8 \pm 3.9$ & $20.6 \pm 4.0$ &
	$16.3 \pm 3.9$ & $17.2 \pm 3.9$
\\
${\rm H}\epsilon + {\rm He\;I}\;3970\,$\tablenotemark{a} & $+0.269$ &
	$20.8 \pm 2.2$ & $26.6 \pm 2.7$ &
	$22.2 \pm 3.9$ & $23.0 \pm 4.0$ &
	$20.5 \pm 4.0$ & $21.0 \pm 4.0$
\\
${\rm H}\delta\;4101$ & $+0.232$ &
	$18.6 \pm 2.4$ & $23.1 \pm 2.9$ &
	$27.1 \pm 2.2$ & $27.6 \pm 2.3$ &
	$29.0 \pm 2.2$ & $29.6 \pm 2.2$
\\
${\rm H}\gamma\;4340$ & $+0.158$ &
	$42.6 \pm 1.6$ & $48.7 \pm 1.8$ &
	$44.6 \pm 2.2$ & $45.3 \pm 2.3$ &
	$42.7 \pm 1.7$ & $43.0 \pm 1.7$
\\
$[{\rm O\;III}]\;4363$ & $+0.151$ &
	\nodata & \nodata & 
	\nodata & \nodata &
	$5.2 \pm 1.4$ & $5.2 \pm 1.4$
\\
${\rm H}\beta\;4861$ & \phs0.000 &
	$100.0 \pm 1.7$ & $100.0 \pm 1.7$ &
	$100.0 \pm 2.6$ & $100.0 \pm 2.6$ &
	$100.0 \pm 2.0$ & $100.0 \pm 2.0$
\\
$[{\rm O\;III}]\;4959$ & $-0.026$ &
	$26.3 \pm 1.4$ & $25.6 \pm 1.4$ &
	$126.9 \pm 3.0$ & $126.4 \pm 3.0$ &
	$123.7 \pm 6.3$ & $122.7 \pm 6.3$
\\
$[{\rm O\;III}]\;5007$ & $-0.038$ &
	$76.8 \pm 1.8$ & $74.0 \pm 1.7$ &
	$378.1 \pm 3.7$ & $376.1 \pm 3.7$ &
	$368.6 \pm 7.8$ & $365.6 \pm 7.7$
\\[1mm]
\multicolumn{2}{c}{$F(\hbeta)$ (ergs s$^{-1}$ cm$^{-2}$)} & 
        \multicolumn{2}{c}{$(1.279 \pm 0.022) \times 10^{-15}$} &
        \multicolumn{2}{c}{$(9.28 \pm 0.24) \times 10^{-15}$} &
        \multicolumn{2}{c}{$(2.985 \pm 0.058) \times 10^{-15}$} 
\\
\multicolumn{2}{c}{EW$_{\rm e}$(\hbeta) (\AA)} &
        \multicolumn{2}{c}{$350 \pm 50$} &
        \multicolumn{2}{c}{$920 \pm 500$} &
        \multicolumn{2}{c}{$242 \pm 26$} 
\\
\multicolumn{2}{c}{Derived $E(B-V)$ (mag)$\,$\tablenotemark{b}} &
        \multicolumn{2}{c}{$+0.18 \pm 0.16$} &
        \multicolumn{2}{c}{$+0.09 \pm 0.20$} &
        \multicolumn{2}{c}{$+0.16 \pm 0.17$} 
\\
\multicolumn{2}{c}{$c(\hbeta)$} &
        \multicolumn{2}{c}{0.36} &
        \multicolumn{2}{c}{0.035} &
        \multicolumn{2}{c}{0} 
\\
\multicolumn{2}{c}{Adopted $A_V$ (mag)} &
        \multicolumn{2}{c}{$+0.77$} &
        \multicolumn{2}{c}{$+0.07$} &
        \multicolumn{2}{c}{0} 
\\
\multicolumn{2}{c}{EW$_{\rm abs}$ (\AA)} &
        \multicolumn{2}{c}{2} &
        \multicolumn{2}{c}{2} &
        \multicolumn{2}{c}{2} 
\\[1mm]
\tableline
& & \multicolumn{2}{c}{Hu III} & 
\multicolumn{2}{c}{Hu III} &
\multicolumn{2}{c}{Hu III} \\
& & \multicolumn{2}{c}{SE ring} & 
\multicolumn{2}{c}{NW arc} & 
\multicolumn{2}{c}{NW ring} \\ 
\multicolumn{1}{c}{Property} &
\multicolumn{1}{c}{$f(\lambda)$} &
\multicolumn{1}{c}{$F$} & \multicolumn{1}{c}{$I$} &
\multicolumn{1}{c}{$F$} & \multicolumn{1}{c}{$I$} &
\multicolumn{1}{c}{$F$} & \multicolumn{1}{c}{$I$} \\
\tableline
$[\rm{O\;II}]\;3727$ & $+0.325$ &
	$289.0 \pm 7.0$ & $347.0 \pm 8.4$ &
	$325.8 \pm 9.8$ & $366 \pm 11$ &
	$352.1 \pm 7.6$ & $323.4 \pm 7.6$
\\
$[{\rm Ne\;III}]\;3869$ & $+0.294$ &
	$27.9 \pm 2.7$ & $32.9 \pm 3.2$ &
	\nodata & \nodata &
	\nodata & \nodata
\\
${\rm H}8 + {\rm He\;I}\;3889$ & $+0.289$ &
	$24.2 \pm 2.3$ & $30.1 \pm 2.7$ &
	\nodata & \nodata &
	\nodata & \nodata
\\
${\rm H}\epsilon + {\rm He\;I}\;3970\,$\tablenotemark{a} & $+0.269$ &
	$18.0 \pm 2.5$ & $22.2 \pm 2.9$ &
	\nodata & \nodata &
	\nodata & \nodata
\\
${\rm H}\delta\;4101$ & $+0.232$ &
        $28.1 \pm 2.1$ & $32.7 \pm 2.4$ & 
	$32.1 \pm 3.1$ & $34.9 \pm 3.4$ &
	$26.4 \pm 2.4$ & $26.8 \pm 2.4$
\\
${\rm H}\gamma\;4340$ & $+0.158$ &
	$42.3 \pm 2.2$ & $46.9 \pm 2.4$ &
	$44.0 \pm 2.9$ & $46.7 \pm 3.1$ &
	$50.5 \pm 2.8$ & $50.5 \pm 2.8$
\\
${\rm H}\beta\;4861$ & \phs0.000 &
	$100.0 \pm 2.2$ & $100.0 \pm 2.2$ &
	$100.0 \pm 4.4$ & $100.0 \pm 4.4$ &
	$100.0 \pm 2.5$ & $100.0 \pm 2.5$
\\
$[{\rm O\;III}]\;4959$ & $-0.026$ &
	$103.6 \pm 2.1$ & $101.6 \pm 2.1$ &
	$60.5 \pm 4.2$ & $59.7 \pm 4.1$ &
	$49.9 \pm 2.1$ & $49.6 \pm 2.1$
\\
$[{\rm O\;III}]\;5007$ & $-0.038$ &
	$299.6 \pm 2.6$ & $291.9 \pm 2.5$ &
	$190.1 \pm 5.1$ & $186.7 \pm 5.0$ &
	$144.9 \pm 2.5$ & $144.1 \pm 2.5$
\\[1mm]
\multicolumn{2}{c}{$F(\hbeta)$ (ergs s$^{-1}$ cm$^{-2}$)} & 
        \multicolumn{2}{c}{$(5.61 \pm 0.12) \times 10^{-15}$} &
        \multicolumn{2}{c}{$(1.817 \pm 0.080) \times 10^{-15}$} &
        \multicolumn{2}{c}{$(2.718 \pm 0.069) \times 10^{-15}$} 
\\
\multicolumn{2}{c}{EW$_{\rm e}$(\hbeta) (\AA)} &
        \multicolumn{2}{c}{$470 \pm 110$} &
        \multicolumn{2}{c}{$530 \pm 300$} &
        \multicolumn{2}{c}{$375 \pm 80$} 
\\
\multicolumn{2}{c}{Derived $E(B-V)$ (mag)$\,$\tablenotemark{b}} &
        \multicolumn{2}{c}{$+0.18 \pm 0.19$} &
        \multicolumn{2}{c}{$+0.12 \pm 0.26$} &
        \multicolumn{2}{c}{$-0.14 \pm 0.21$} 
\\
\multicolumn{2}{c}{$c(\hbeta)$} &
        \multicolumn{2}{c}{0.25} &
        \multicolumn{2}{c}{0.16} &
        \multicolumn{2}{c}{0} 
\\
\multicolumn{2}{c}{Adopted $A_V$ (mag)} &
        \multicolumn{2}{c}{$+0.53$} &
        \multicolumn{2}{c}{$+0.34$} &
        \multicolumn{2}{c}{0} 
\\
\multicolumn{2}{c}{EW$_{\rm abs}$ (\AA)} &
        \multicolumn{2}{c}{2} &
        \multicolumn{2}{c}{2} &
        \multicolumn{2}{c}{2} 
\\[1mm]
\tableline
\end{tabular}
\vspace*{2mm}
\tablenotetext{a}{
Blended with [\ion{Ne}{3}]$\lambda$ 3967.
}
\tablenotetext{b}{
Derived from $F$(\hgamma)/$F$(\hbeta).
}
\tablecomments{
Emission lines are listed in \AA.
$F$ is the observed flux ratio with respect to \hbeta.
$I$ is the corrected intensity ratio, corrected for the adopted
reddening listed, and for underlying Balmer absorption.
The uncertainties in the observed line ratios account for the
uncertainties in the fits to the line profiles, the surrounding
continua, and the relative uncertainty in the sensitivity function
listed in Table~\ref{table_obslog}. 
Flux uncertainties in the \hbeta\ reference line are not included.
Uncertainties in the corrected line ratios account for uncertainties
in the specified line and in the \hbeta\ reference line.
The reddening function, $f(\lambda)$, is given.
Also listed are the observed \hbeta\ flux; the equivalent width of
\hbeta\ in emission, EW$_{\rm e}$(\hbeta).
Where \othreea\ is measured, simultaneous solutions for the
logarithmic reddening, $c(\hbeta)$, and
the equivalent width of the underlying Balmer absorption,
EW$_{\rm abs}$ are listed.
The adopted value of the extinction in $V$, $A_V$, is listed.
Where \othreea\ is not measured, the equivalent width of the
underlying Balmer absorption was set to 2~\AA.
}
\end{center}
\end{table}

\begin{table}
\setlength{\tabcolsep}{2mm}
\tablenum{4b}
\begin{center}
\renewcommand{\arraystretch}{1.0} 
\caption{
Blue spectra: line ratios and properties for nebulae
Hubble~V, K$\alpha$, K$\beta$, KD~22e, and KD~28e.
\vspace*{3mm}
\label{table_gr07data2}
}
\begin{tabular}{rccccccc}
\tableline \tableline
& & \multicolumn{2}{c}{Hu V} & 
\multicolumn{2}{c}{K$\alpha$} & 
\multicolumn{2}{c}{K$\beta$} \\ 
\multicolumn{1}{c}{Property} &
\multicolumn{1}{c}{$f(\lambda)$} &
\multicolumn{1}{c}{$F$} & \multicolumn{1}{c}{$I$} &
\multicolumn{1}{c}{$F$} & \multicolumn{1}{c}{$I$} &
\multicolumn{1}{c}{$F$} & \multicolumn{1}{c}{$I$} \\
\tableline
$[\rm{O\;II}]\;3727$ & $+0.325$ &
	$115.98 \pm 0.91$ & $141.0 \pm 1.1$ &
	$79.95 \pm 0.94$ & $97.9 \pm 1.2$ &
	$154.8 \pm 2.2$ & $202.2 \pm 2.9$ 
\\
${\rm H}12\;3751$ & $+0.320$ &
	$1.84 \pm 0.70$ & $2.79 \pm 0.85$ &
	\nodata & \nodata &
	$1.6 \pm 1.7$ & $2.4 \pm 2.2$
\\
${\rm H}11\;3772$ & $+0.316$ &
	$2.62 \pm 0.72$ & $3.84 \pm 0.87$ &
	$1.89 \pm 0.71$ & $3.16 \pm 0.86$ &
	$2.4 \pm 1.7$ & $3.4 \pm 2.2$
\\
${\rm H}10\;3799$ & $+0.310$ &
	$4.14 \pm 0.76$ & $5.83 \pm 0.91$ &
	$3.30 \pm 0.75$ & $5.07 \pm 0.91$ &
	$3.8 \pm 1.8$ & $5.2 \pm 2.3$
\\
${\rm H}9\;3835$ & $+0.302$ &
	$5.09 \pm 0.63$ & $7.15 \pm 0.75$ &
	$5.83 \pm 0.65$ & $8.21 \pm 0.78$ &
	$7.9 \pm 1.3$ & $10.5 \pm 1.7$
\\
$[{\rm Ne\;III}]\;3869$ & $+0.294$ &
	$30.81 \pm 0.74$ & $36.73 \pm 0.88$ &
	$39.93 \pm 0.76$ & $47.93 \pm 0.91$ &
	$35.5 \pm 1.4$ & $45.2 \pm 1.8$ 
\\
${\rm H}8 + {\rm He\;I}\;3889$ & $+0.289$ &
	$14.85 \pm 0.65$ & $18.91 \pm 0.77$ &
	$15.00 \pm 0.65$ & $19.35 \pm 0.78$ &
	$18.4 \pm 1.2$ & $23.8 \pm 1.5$ 
\\
${\rm H}\epsilon + {\rm He\;I}\;3970\,$\tablenotemark{a} & $+0.269$ &
	$21.00 \pm 0.36$ & $25.87 \pm 0.42$ &
	$21.87 \pm 0.37$ & $27.17 \pm 0.44$ &
	$20.4 \pm 1.3$ & $25.9 \pm 1.6$
\\
${\rm He\;I}\;4027$ & $+0.253$ &
	$1.64 \pm 0.31$ & $1.91 \pm 0.36$ &
	$1.35 \pm 0.30$ & $1.58 \pm 0.35$ &
	$1.21 \pm 0.48$ & $1.49 \pm 0.59$
\\
$[{\rm S\;II}]\;4068,4076$ & $+0.241$ &
	$1.02 \pm 0.32$ & $1.18 \pm 0.37$ &
	$0.64 \pm 0.35$ & $0.74 \pm 0.41$ &
	$1.32 \pm 0.46$ & $1.61 \pm 0.56$
\\
${\rm H}\delta\;4101$ & $+0.232$ &
	$21.70 \pm 0.37$ & $26.05 \pm 0.42$ &
	$21.88 \pm 0.44$ & $26.45 \pm 0.51$ &
	$21.27 \pm 0.58$ & $26.10 \pm 0.70$
\\
${\rm H}\gamma\;4340$ & $+0.158$ &
	$41.63 \pm 0.78$ & $46.53 \pm 0.85$ &
	$40.9 \pm 1.0$ & $46.0 \pm 1.1$ &
	$40.6 \pm 1.3$ & $46.5 \pm 1.5$
\\
$[{\rm O\;III}]\;4363$ & $+0.151$ &
	$5.22 \pm 0.60$ & $5.69 \pm 0.65$ &
	$6.75 \pm 0.81$ & $7.39 \pm 0.89$ &
	$4.9 \pm 1.0$ & $5.5 \pm 1.1$
\\
${\rm He\;I}\;4388$ & $+0.143$ &
	$0.47 \pm 0.60$ & $0.51 \pm 0.65$ &
	\nodata & \nodata &
	\nodata & \nodata
\\
${\rm He\;I}\;4471$ & $+0.116$ &
	$3.48 \pm 0.11$ & $3.71 \pm 0.12$ &
	$2.84 \pm 0.23$ & $3.04 \pm 0.25$ &
	$3.55 \pm 0.48$ & $3.90 \pm 0.53$
\\
$[{\rm Ar\;IV}] + {\rm He\;I}\;4713$ & $+0.042$ &
	$0.61 \pm 0.07$ & $0.62 \pm 0.07$ &
	\nodata & \nodata &
	\nodata & \nodata
\\
${\rm H}\beta\;4861$ & \phs0.000 &
	$100.0 \pm 1.3$ & $100.0 \pm 1.3$ &
	$100.0 \pm 1.0$ & $100.00 \pm 0.99$ &
	$100.0 \pm 1.2$ & $100.0 \pm 1.2$
\\
${\rm He\;I}\;4922$ & $-0.016$ &
	$0.97 \pm 0.09$ & $0.95 \pm 0.08$ &
	$2.1 \pm 3.7$ & $2.1 \pm 3.6$ &
	\nodata & \nodata
\\
$[{\rm O\;III}]\;4959$ & $-0.026$ &
	$179.7 \pm 4.3$ & $175.6 \pm 4.2$ &
	$198.3 \pm 3.7$ & $193.7 \pm 3.6$ &
	$177.7 \pm 3.3$ & $173.5 \pm 3.2$
\\
$[{\rm O\;III}]\;5007$ & $-0.038$ &
	$534.8 \pm 5.3$ & $518.6 \pm 5.1$ &
	$592.8 \pm 4.5$ & $574.4 \pm 4.4$ &
	$533.8 \pm 3.9$ & $515.8 \pm 3.8$
\\[1mm]
\multicolumn{2}{c}{$F(\hbeta)$ (ergs s$^{-1}$ cm$^{-2}$)} & 
        \multicolumn{2}{c}{$(9.48 \pm 0.12) \times 10^{-14}$} &
        \multicolumn{2}{c}{$(1.328 \pm 0.014) \times 10^{-14}$} &
        \multicolumn{2}{c}{$(6.438 \pm 0.079) \times 10^{-15}$} 
\\
\multicolumn{2}{c}{EW$_{\rm e}$(\hbeta) (\AA)} &
        \multicolumn{2}{c}{$239 \pm 18$} &
        \multicolumn{2}{c}{$261 \pm 18$} &
        \multicolumn{2}{c}{$507 \pm 77$} 
\\
\multicolumn{2}{c}{Derived $E(B-V)$ (mag)$\,$\tablenotemark{b}} &
        \multicolumn{2}{c}{$+0.20 \pm 0.14$} &
        \multicolumn{2}{c}{$+0.23 \pm 0.14$} &
        \multicolumn{2}{c}{$+0.26 \pm 0.15$} 
\\
\multicolumn{2}{c}{$c(\hbeta)$} &
        \multicolumn{2}{c}{0.27} &
        \multicolumn{2}{c}{0.28} &
        \multicolumn{2}{c}{0.36}
\\
\multicolumn{2}{c}{Adopted $A_V$ (mag)} &
        \multicolumn{2}{c}{$+0.57$} &
        \multicolumn{2}{c}{$+0.60$} &
        \multicolumn{2}{c}{$+0.77$} 
\\
\multicolumn{2}{c}{EW$_{\rm abs}$ (\AA)} &
        \multicolumn{2}{c}{1.7} &
        \multicolumn{2}{c}{1.8} &
        \multicolumn{2}{c}{1.3} 
\\[1mm]
\tableline
& & \multicolumn{2}{c}{KD 22e} & 
\multicolumn{2}{c}{KD 28e} & \\ 
\multicolumn{1}{c}{Property} &
\multicolumn{1}{c}{$f(\lambda)$} &
\multicolumn{1}{c}{$F$} & \multicolumn{1}{c}{$I$} &
\multicolumn{1}{c}{$F$} & \multicolumn{1}{c}{$I$} \\
\tableline
$[\rm{O\;II}]\;3727$ & $+0.325$ &
	$478.2 \pm 6.8$ & $472.8 \pm 6.7$ &
	$218.6 \pm 1.9$ & $239.6 \pm 2.1$ &
	&
\\
${\rm H}12\;3751$ & $+0.320$ &
	\nodata & \nodata &
	$1.7 \pm 1.5$ & $2.3 \pm 1.6$ &
	&
\\
${\rm H}11\;3772$ & $+0.316$ &
	\nodata & \nodata &
	$5.3 \pm 1.5$ & $6.2 \pm 1.6$ &
	&
\\
${\rm H}10\;3799$ & $+0.310$ &
	\nodata & \nodata &
	$6.3 \pm 1.6$ & $7.4 \pm 1.8$ &
	&
\\
${\rm H}9\;3835$ & $+0.302$ &
	\nodata & \nodata &
	$6.32 \pm 0.76$ & $7.74 \pm 0.83$ &
	&
\\
$[{\rm Ne\;III}]\;3869$ & $+0.294$ &
	\nodata & \nodata &
	$26.97 \pm 0.88$ & $29.29 \pm 0.96$ &
	&
\\
${\rm H}8 + {\rm He\;I}\;3889$ & $+0.289$ &
	\nodata & \nodata &
	$17.59 \pm 0.82$ & $20.01 \pm 0.89$ &
	&
\\
${\rm H}\epsilon + {\rm He\;I}\;3970\,$\tablenotemark{a} & $+0.269$ &
	\nodata & \nodata &
	$21.15 \pm 0.61$ & $23.72 \pm 0.66$ &
	&
\\
${\rm H}\delta\;4101$ & $+0.232$ &
	$24.2 \pm 3.4$ & $26.3 \pm 3.4$ &
	$23.43 \pm 0.62$ & $25.85 \pm 0.66$ &
	&
\\
${\rm H}\gamma\;4340$ & $+0.158$ &
	$51.0 \pm 2.1$ & $52.2 \pm 2.1$ &
	$44.34 \pm 0.74$ & $46.88 \pm 0.77$ &
	&
\\
$[{\rm O\;III}]\;4363$ & $+0.151$ &
	\nodata & \nodata &
	$3.99 \pm 0.59$ & $4.15 \pm 0.61$ &
	&
\\
${\rm He\;I}\;4471$ & $+0.116$ &
	\nodata & \nodata &
	$2.74 \pm 0.29$ & $2.82 \pm 0.30$ &
	&
\\
${\rm H}\beta\;4861$ & \phs0.000 &
	$100.0 \pm 2.6$ & $100.0 \pm 2.6$ &
	$100.00 \pm 0.79$ & $100.00 \pm 0.79$ &
	&
\\
$[{\rm O\;III}]\;4959$ & $-0.026$ &
	$28.7 \pm 2.2$ & $28.4 \pm 2.2$ &
	$141.8 \pm 1.8$ & $139.9 \pm 1.8$ &
	&
\\
$[{\rm O\;III}]\;5007$ & $-0.038$ &
	$68.9 \pm 2.4$ & $68.1 \pm 2.4$ &
	$424.3 \pm 2.2$ & $417.2 \pm 2.2$ &
	&
\\[1mm]
\multicolumn{2}{c}{$F(\hbeta)$ (ergs s$^{-1}$ cm$^{-2}$)} & 
        \multicolumn{2}{c}{$(3.347 \pm 0.086) \times 10^{-15}$} &
        \multicolumn{2}{c}{$(4.276 \pm 0.034) \times 10^{-15}$} 
\\
\multicolumn{2}{c}{EW$_{\rm e}$(\hbeta) (\AA)} &
        \multicolumn{2}{c}{$174 \pm 18$} &
        \multicolumn{2}{c}{$396 \pm 32$} 
\\
\multicolumn{2}{c}{Derived $E(B-V)$ (mag)$\,$\tablenotemark{b}} &
        \multicolumn{2}{c}{0} &
        \multicolumn{2}{c}{$+0.09 \pm 0.13$} 
\\
\multicolumn{2}{c}{$c(\hbeta)$} &
        \multicolumn{2}{c}{0} &
        \multicolumn{2}{c}{0.13} 
\\
\multicolumn{2}{c}{Adopted $A_V$ (mag)} &
        \multicolumn{2}{c}{0} &
        \multicolumn{2}{c}{$+0.28$} 
\\
\multicolumn{2}{c}{EW$_{\rm abs}$ (\AA)} &
        \multicolumn{2}{c}{2} &
        \multicolumn{2}{c}{2.2} 
\\[1mm]
\tableline
\end{tabular}
\vspace*{2mm}
\tablenotetext{a}{
Blended with [\ion{Ne}{3}]$\lambda$ 3967.
}
\tablenotetext{b}{
Derived from $F$(\hgamma)/$F$(\hbeta).
}
\tablecomments{
See Table~\ref{table_gr07data1} for additional comments.
}
\end{center}
\end{table}

\begin{table}
\setlength{\tabcolsep}{2mm}
\tablenum{5a}
\begin{center}
\renewcommand{\arraystretch}{0.90} 
\caption{
Low-dispersion spectra: line ratios and properties for nebulae
HK~16, HK~42, HK~69, HK~70, Hodge~7, and Hodge~12.
\vspace*{3mm}
\label{table_gr11data1}
}
\begin{tabular}{rccccccc}
\tableline \tableline
& & \multicolumn{2}{c}{HK 16} & 
\multicolumn{2}{c}{HK 42} & 
\multicolumn{2}{c}{HK 69} \\ 
\multicolumn{1}{c}{Property} &
\multicolumn{1}{c}{$f(\lambda)$} &
\multicolumn{1}{c}{$F$} & \multicolumn{1}{c}{$I$} &
\multicolumn{1}{c}{$F$} & \multicolumn{1}{c}{$I$} &
\multicolumn{2}{c}{$F$} \\
\tableline
$[\rm{O\;II}]\;3727$ & $+0.325$ &
	$399 \pm 26$ & $496 \pm 32$ &
	$391 \pm 38$ & $559 \pm 54$ &
	\multicolumn{2}{c}{$424 \pm 19$}
\\
${\rm H}\beta\;4861$ & \phs0.000 &
	$100 \pm 14$ & $100 \pm 13$ &
	$100 \pm 13$ & $100 \pm 12$ &
	\multicolumn{2}{c}{$100 \pm 11$}
\\
$[{\rm O\;III}]\;4959$ & $-0.026$ &
	$6 \pm 10$ & $6 \pm 10$ &
	$126 \pm 16$ & $112 \pm 14$ &
	\multicolumn{2}{c}{\nodata}
\\
$[{\rm O\;III}]\;5007$ & $-0.038$ &
	$62 \pm 11$ & $59 \pm 10$ &
	$353 \pm 21$ & $308 \pm 18$ &
	\multicolumn{2}{c}{\nodata}
\\
${\rm H}\alpha\;6563$ & $-0.299$ &
	$363 \pm 16$ & $285 \pm 12$ &
	$460 \pm 24$ & $286 \pm 15$ &
	\multicolumn{2}{c}{$504 \pm 18$}
\\
$[{\rm N\;II}]\;6583$ & $-0.301$ &
	$44 \pm 13$ & $35 \pm 10$ &
	$17 \pm 19$ & $10 \pm 12$ &
	\multicolumn{2}{c}{$14 \pm 15$}
\\
$[{\rm S\;II}]\;6716$ & $-0.319$ &
	$91 \pm 12$ & $70.3 \pm 9.1$ &
	$10 \pm 18$ & $6 \pm 10$ &
	\multicolumn{2}{c}{$132 \pm 13\,$\tablenotemark{a}}
\\
$[{\rm S\;II}]\;6731$ & $-0.321$ &
	$41 \pm 12$ & $31.6 \pm 9.3$ &
	$31 \pm 18$ & $19 \pm 11$ &
	\multicolumn{2}{c}{$132 \pm 13\,$\tablenotemark{a}}
\\[1mm]
\multicolumn{2}{c}{$F(\hbeta)$ (ergs s$^{-1}$ cm$^{-2}$)} & 
        \multicolumn{2}{c}{$(1.06 \pm 0.15) \times 10^{-16}$} &
        \multicolumn{2}{c}{$(6.88 \pm 0.93) \times 10^{-17}$} &
        \multicolumn{2}{c}{$(1.66 \pm 0.18) \times 10^{-16}$} 
\\
\multicolumn{2}{c}{EW$_{\rm e}$(\hbeta) (\AA)} &
        \multicolumn{2}{c}{$85 \pm 19$} &
        \multicolumn{2}{c}{$23.2 \pm 3.5$} &
        \multicolumn{2}{c}{\nodata\tablenotemark{b}} 
\\
\multicolumn{2}{c}{Derived $E(B-V)$ (mag)$\,$\tablenotemark{c}} &
        \multicolumn{2}{c}{$+0.22 \pm 0.20$} &
        \multicolumn{2}{c}{$+0.41 \pm 0.22$} &
        \multicolumn{2}{c}{$+0.57 \pm 0.16$} 
\\
\multicolumn{2}{c}{$c(\hbeta)$} &
        \multicolumn{2}{c}{0.32} &
        \multicolumn{2}{c}{0.59} &
        \multicolumn{2}{c}{\nodata} 
\\
\multicolumn{2}{c}{Adopted $A_V$ (mag)} &
        \multicolumn{2}{c}{$+0.68$} &
        \multicolumn{2}{c}{$+1.25$} &
        \multicolumn{2}{c}{$+1.74$} 
\\
\multicolumn{2}{c}{EW$_{\rm abs}$ (\AA)} &
        \multicolumn{2}{c}{2} &
        \multicolumn{2}{c}{2} &
        \multicolumn{2}{c}{\nodata} 
\\[1mm]
\tableline
& & \multicolumn{2}{c}{HK 70} & 
\multicolumn{2}{c}{Ho 7} &      
\multicolumn{2}{c}{Ho 12} \\	
\multicolumn{1}{c}{Property} &
\multicolumn{1}{c}{$f(\lambda)$} &
\multicolumn{1}{c}{$F$} & \multicolumn{1}{c}{$I$} &
\multicolumn{1}{c}{$F$} & \multicolumn{1}{c}{$I$} &
\multicolumn{1}{c}{$F$} & \multicolumn{1}{c}{$I$} \\
\tableline
$[\rm{O\;II}]\;3727$ & $+0.325$ &
	$311 \pm 25$ & $488 \pm 39$ &
	$290.2 \pm 9.5$ & $474 \pm 130$ &
	$362.4 \pm 7.3$ & $535 \pm 24$ 
\\
$[{\rm Ne\;III}]\;3869$ & $+0.294$ &
	$60 \pm 10$ & $90 \pm 15$ &
	\nodata & \nodata &
	$19.7 \pm 2.1$ & $28.0 \pm 3.2$
\\
${\rm H}8 + {\rm He\;I}\;3889$ & $+0.289$ &
	\nodata & \nodata &
	\nodata & \nodata &
	$18.8 \pm 2.9$ & $28.0 \pm 4.2$
\\
${\rm H}\epsilon + {\rm He\;I}\;3970\,$\tablenotemark{d} & $+0.269$ &
	\nodata & \nodata &
	\nodata & \nodata &
	$17.7 \pm 1.7$ & $26.2 \pm 2.5$
\\
$[{\rm S\;II}]\;4068,4076$ & $+0.241$ &
	\nodata & \nodata &
	\nodata & \nodata &
	$16.3 \pm 1.8$ & $21.7 \pm 2.5$
\\
${\rm H}\delta\;4101$ & $+0.232$ &
	$33.7 \pm 9.8$ & $46 \pm 14$ &
	\nodata & \nodata &
	$19.8 \pm 1.9$ & $27.4 \pm 2.6$
\\
${\rm H}\gamma\;4340$ & $+0.158$ &
	$48.8 \pm 8.5$ & $61 \pm 11$ &
	$27.4 \pm 2.8$ & $33.3 \pm 8.2$ &
	$40.8 \pm 2.0$ & $50.3 \pm 2.6$
\\
$[{\rm O\;III}]\;4363$ & $+0.151$ &
	\nodata & \nodata &
	\nodata & \nodata &
	$5.4 \pm 1.6$ & $6.4 \pm 1.9$
\\
${\rm H}\beta\;4861$ & \phs0.000 &
	$100 \pm 12$ & $100 \pm 12$ &
	$100.0 \pm 6.0$ & $100.0 \pm 6.9$ &
	$100.0 \pm 3.7$ & $100.0 \pm 3.7$
\\
$[{\rm O\;III}]\;4959$ & $-0.026$ &
	$152 \pm 13$ & $147 \pm 13$ &
	$72.5 \pm 6.3$ & $71 \pm 11$ &
	$21.1 \pm 1.6$ & $20.3 \pm 1.5$
\\
$[{\rm O\;III}]\;5007$ & $-0.038$ &
	$463 \pm 17$ & $440 \pm 16$ &
	$203.8 \pm 8.2$ & $197 \pm 21$ &
	$59.8 \pm 2.0$ & $56.6 \pm 1.9$
\\
${\rm He\;I}\;5876$ & $-0.204$ &
	\nodata & \nodata &
	\nodata & \nodata &
	$10.5 \pm 1.6$ & $8.1 \pm 1.3$
\\
$[{\rm O\;I}]\;6300 + [{\rm S\;III}]\;6312$ & $-0.264$ &
	\nodata & \nodata &
	\nodata & \nodata &
	$73.6 \pm 4.8$ & $52.8 \pm 3.8$
\\
$[{\rm O\;I}]\;6363$ & $-0.272$ &
	\nodata & \nodata &
	\nodata & \nodata &
	$27.0 \pm 2.1$ & $19.2 \pm 1.6$
\\
${\rm H}\alpha\;6563$ & $-0.299$ &
	$442 \pm 15$ & $292.0 \pm 9.7$ &
	$403 \pm 13$ & $286 \pm 88$ &
	$411.9 \pm 9.3$ & $284 \pm 12$
\\
$[{\rm N\;II}]\;6583$ & $-0.301$ &
	$33 \pm 12$ & $21.7 \pm 8.0$ &
	$4 \pm 11$ & $3 \pm 8$ &
	$38.8 \pm 7.8$ & $26.6 \pm 5.4$
\\
$[{\rm S\;II}]\;6716$ & $-0.319$ &
	$86 \pm 11$ & $55.5 \pm 7.3$ &
	\nodata & \nodata &
	$226.2 \pm 5.3\,$\tablenotemark{a} & $151.7 \pm 6.9$
\\
$[{\rm S\;II}]\;6731$ & $-0.321$ &
	$52 \pm 12$ & $33.4 \pm 7.4$ &
	\nodata & \nodata &
	$226.2 \pm 5.3\,$\tablenotemark{a} & $151.7 \pm 6.9$
\\
${\rm He\;I}\;7281$ & $-0.395$ &
        \nodata & \nodata &
 	\nodata & \nodata &
 	$34.5 \pm 4.1$ & $21.1 \pm 2.7$
\\
$[{\rm O\;II}]\;7320,7330$ & $-0.400$ &
	\nodata & \nodata &
	\nodata & \nodata &
	$46.1 \pm 4.4$ & $28.0 \pm 3.0$
\\[1mm]
\multicolumn{2}{c}{$F(\hbeta)$ (ergs s$^{-1}$ cm$^{-2}$)} & 
        \multicolumn{2}{c}{$(1.36 \pm 0.17) \times 10^{-16}$} &
        \multicolumn{2}{c}{$(1.233 \pm 0.074) \times 10^{-16}$} &
        \multicolumn{2}{c}{$(1.084 \pm 0.040) \times 10^{-15}$} 
\\
\multicolumn{2}{c}{EW$_{\rm e}$(\hbeta) (\AA)} &
        \multicolumn{2}{c}{$74 \pm 15$} &
        \multicolumn{2}{c}{\nodata\tablenotemark{b}} &
        \multicolumn{2}{c}{$148 \pm 16$} 
\\
\multicolumn{2}{c}{Derived $E(B-V)$ (mag)$\,$\tablenotemark{c}} &
        \multicolumn{2}{c}{$+0.42 \pm 0.18$} &
        \multicolumn{2}{c}{$+0.35 \pm 0.11$} &
        \multicolumn{2}{c}{$+0.359 \pm 0.089$} 
\\
\multicolumn{2}{c}{$c(\hbeta)$} &
        \multicolumn{2}{c}{0.60} &
        \multicolumn{2}{c}{\nodata} &
        \multicolumn{2}{c}{$0.530 \pm 0.053$} 
\\
\multicolumn{2}{c}{Adopted $A_V$ (mag)} &
        \multicolumn{2}{c}{$+1.29$} &
        \multicolumn{2}{c}{$+1.07$} &
        \multicolumn{2}{c}{$+1.13$} 
\\
\multicolumn{2}{c}{EW$_{\rm abs}$ (\AA)} &
        \multicolumn{2}{c}{0} &
        \multicolumn{2}{c}{2} &
        \multicolumn{2}{c}{1.3} 
\\[1mm]
\tableline
\end{tabular}
\vspace*{2mm}
\tablenotetext{a}{
[\ion{S}{2}] unresolved.
}
\tablenotetext{b}{
Very weak continuum.
}
\tablenotetext{c}{
Derived from $F$(\halpha)/$F$(\hbeta).
}
\tablenotetext{d}{
Blended with [\ion{Ne}{3}]$\lambda$ 3967.
}
\tablecomments{
See Table~\ref{table_gr07data1} for additional comments.
}
\end{center}
\end{table}

\begin{table}
\scriptsize 
\setlength{\tabcolsep}{2mm}
\tablenum{5b}
\begin{center}
\renewcommand{\arraystretch}{1.} 
\caption{
Low-dispersion spectra: line ratios and properties for nebulae
Hubble~V, K$\alpha$, K$\beta$, KD~9, KD~20, and KD~21.
\vspace*{3mm}
\label{table_gr11data2}
}
\begin{tabular}{rccccccc}
\tableline \tableline
& & \multicolumn{2}{c}{Hu V} &  
\multicolumn{2}{c}{K$\alpha$} & 
\multicolumn{2}{c}{K$\beta$} \\ 
\multicolumn{1}{c}{Property} &
\multicolumn{1}{c}{$f(\lambda)$} &
\multicolumn{1}{c}{$F$} & \multicolumn{1}{c}{$I$} &
\multicolumn{1}{c}{$F$} & \multicolumn{1}{c}{$I$} &
\multicolumn{1}{c}{$F$} & \multicolumn{1}{c}{$I$} \\
\tableline
$[\rm{O\;II}]\;3727$ & $+0.325$ &
	$90.2 \pm 1.6$ & $128.9 \pm 6.4$ &
	$76.9 \pm 1.5$ & $125.4 \pm 5.2$ &
	$404 \pm 24$ & $574 \pm 84$ 
\\
${\rm H}12\;3751$ & $+0.320$ &
	\nodata & \nodata &
	$1.2 \pm 1.1$ & $1.9 \pm 1.8$ &
	\nodata & \nodata
\\
${\rm H}11\;3772$ & $+0.316$ &
	$2.0 \pm 1.2$ & $2.8 \pm 1.7$ &
	$2.6 \pm 1.1$ & $4.2 \pm 1.8$ &
	\nodata & \nodata
\\
${\rm H}10\;3799$ & $+0.310$ &
	$3.3 \pm 1.3$ & $4.6 \pm 1.8$ &
	$4.8 \pm 1.2$ & $7.7 \pm 1.9$ &
	\nodata & \nodata
\\
${\rm H}9\;3835$ & $+0.302$ &
	$5.57 \pm 0.71$ & $7.8 \pm 1.0$ &
	$4.74 \pm 0.80$ & $7.5 \pm 1.3$ &
	\nodata & \nodata
\\
$[{\rm Ne\;III}]\;3869$ & $+0.294$ &
	$35.14 \pm 0.75$ & $48.5 \pm 2.3$ &
	$39.19 \pm 0.93$ & $61.0 \pm 2.5$ &
	\nodata & \nodata
\\
${\rm H}8 + {\rm He\;I}\;3889$ & $+0.289$ &
	$14.88 \pm 0.65$ & $20.4 \pm 1.2$ &
	$14.11 \pm 0.80$ & $21.8 \pm 1.4$ &
	\nodata & \nodata
\\
${\rm H}\epsilon + {\rm He\;I}\;3970\,$\tablenotemark{a} & $+0.269$ &
	$23.63 \pm 0.50$ & $31.8 \pm 1.4$ &
	$20.83 \pm 0.60$ & $31.2 \pm 1.3$ &
	\nodata & \nodata
\\
${\rm He\;I}\;4027$ & $+0.253$ &
	$1.36 \pm 0.36$ & $1.80 \pm 0.48$ &
	$1.05 \pm 0.49$ & $1.54 \pm 0.72$ &
	\nodata & \nodata
\\
$[{\rm S\;II}]\;4068,4076$ & $+0.241$ &
	$0.91 \pm 0.49$ & $1.19 \pm 0.64$ &
	$0.61 \pm 0.48$ & $0.88 \pm 0.69$ &
	\nodata & \nodata
\\
${\rm H}\delta\;4101$ & $+0.232$ &
	$21.44 \pm 0.58$ & $27.7 \pm 1.2$ &
	$20.78 \pm 0.57$ & $29.4 \pm 1.1$ &
	\nodata & \nodata 
\\
${\rm H}\gamma\;4340$ & $+0.158$ &
	$40.59 \pm 0.98$ & $48.3 \pm 1.6$ &
	$40.37 \pm 0.98$ & $51.2 \pm 1.5$ &
	$40.4 \pm 6.4$ & $47 \pm 12$
\\
$[{\rm O\;III}]\;4363$ & $+0.151$ &
	$5.05 \pm 0.72$ & $5.96 \pm 0.86$ &
	$7.47 \pm 0.78$ & $9.37 \pm 0.99$ &
	\nodata & \nodata
\\
${\rm He\;I}\;4388$ & $+0.143$ &
	$0.54 \pm 0.74$ & $0.63 \pm 0.87$ &
	\nodata & \nodata &
	\nodata & \nodata
\\
${\rm He\;I}\;4471$ & $+0.116$ &
	$3.62 \pm 0.14$ & $4.11 \pm 0.17$ &
	$3.80 \pm 0.21$ & $4.53 \pm 0.26$ &
	\nodata & \nodata
\\
$[{\rm Ar\;IV}] + {\rm He\;I}\;4713$ & $+0.042$ &
	$0.645 \pm 0.070$ & $0.68 \pm 0.07$ &
	$0.99 \pm 0.15$ & $1.05 \pm 0.16$ &
	\nodata & \nodata
\\
$[{\rm Ar\;IV}]\;4740$ & $+0.034$ &
	$0.214 \pm 0.054$ & $0.22 \pm 0.06$ &
	\nodata & \nodata &
	\nodata & \nodata 
\\
${\rm H}\beta\;4861$ & \phs0.000 &
	$100.0 \pm 3.8$ & $100.0 \pm 3.8$ &
	$100.0 \pm 2.7$ & $100.0 \pm 2.7$ &
	$100.0 \pm 7.4$ & $100.0 \pm 8.1$
\\
${\rm He\;I}\;4922$ & $-0.016$ &
	$1.06 \pm 0.12$ & $1.04 \pm 0.12$ &
	$1.2 \pm 2.3$ & $1.2 \pm 2.2$ &
	\nodata & \nodata
\\
$[{\rm O\;III}]\;4959$ & $-0.026$ &
	$195 \pm 14$ & $189 \pm 13$ &
	$197 \pm 13$ & $189 \pm 12$ &
	$18.0 \pm 5.9$ & $17.7 \pm 7.3$
\\
$[{\rm O\;III}]\;5007$ & $-0.038$ &
	$572 \pm 18$ & $549 \pm 18$ &
	$596 \pm 17$ & $563 \pm 16$ &
	$70.2 \pm 7.0$ & $68 \pm 13$
\\
$[{\rm N\;I}]\;5199$ & $-0.083$ &
	$0.160 \pm 0.043$ & $0.15 \pm 0.04$ &
	$0.47 \pm 0.11$ & $0.41 \pm 0.10$ &
	\nodata & \nodata
\\
${\rm He\;I}\;5876$ & $-0.204$ &
	$14.16 \pm 0.51$ & $11.31 \pm 0.52$ &
	$16.15 \pm 0.58$ & $11.87 \pm 0.51$ &
	\nodata & \nodata
\\
$[{\rm O\;I}]\;6300$ & $-0.264$ &
	$0.11 \pm 0.33$ & $0.08 \pm 0.25$ &
	$3.19 \pm 0.30$ & $2.14 \pm 0.21$ &
	\nodata & \nodata
\\
$[{\rm S\;III}]\;6312$ & $-0.266$ &
	$3.58 \pm 0.33$ & $2.67 \pm 0.26$ &
	$1.33 \pm 0.27$ & $0.89 \pm 0.18$ &
 	\nodata & \nodata
\\
$[{\rm O\;I}]\;6363$ & $-0.272$ &
	$0.58 \pm 0.13$ & $0.43 \pm 0.10$ &
	$0.85 \pm 0.27$ & $0.56 \pm 0.18$ &
	\nodata & \nodata
\\
${\rm H}\alpha\;6563$ & $-0.299$ &
	$405 \pm 10$ & $291 \pm 14$ &
	$457 \pm 10$ & $291 \pm 12$ &
	$369 \pm 12$ & $286 \pm 36$
\\
$[{\rm N\;II}]\;6583$ & $-0.301$ &
        $1.98 \pm 0.12$ & $1.42 \pm 0.11$ & 
	$3.74 \pm 0.46$ & $2.37 \pm 0.30$ & 
	$8 \pm 10$ & $6.3 \pm 8.6$
\\
${\rm He\;I}\;6678$ & $-0.314$ &
	$4.75 \pm 0.34$ & $3.36 \pm 0.28$ &
	$5.39 \pm 0.28$ & $3.36 \pm 0.21$ &
	\nodata & \nodata
\\
$[{\rm S\;II}]\;6716$ & $-0.319$ &
	$8.36 \pm 0.42$ & $5.88 \pm 0.40$ &
	$8.17 \pm 0.28$ & $5.05 \pm 0.25$ &
	$51.6 \pm 8.6\,$\tablenotemark{b} & $39.5 \pm 9.9$
\\
$[{\rm S\;II}]\;6731$ & $-0.321$ &
	$6.18 \pm 0.38$ & $4.34 \pm 0.33$ &
	$7.09 \pm 0.27$ & $4.37 \pm 0.23$ &
	$51.6 \pm 8.6\,$\tablenotemark{b} & $39.5 \pm 9.9$
\\
${\rm He\;I}\;7065$ & $-0.366$ &
	$4.10 \pm 0.18$ & $2.74 \pm 0.19$ &
	$12.01 \pm 0.24$ & $6.92 \pm 0.32$ &
	\nodata & \nodata 
\\
$[{\rm Ar\;III}]\;7136$ & $-0.375$ &
	$14.91 \pm 0.23$ & $9.87 \pm 0.55$ &
	$18.75 \pm 0.26$ & $10.7 \pm 0.47$ &
	\nodata & \nodata
\\
${\rm He\;I}\;7281$ & $-0.395$ &
 	$1.011 \pm 0.082$ & $0.66 \pm 0.06$ &
 	$1.70 \pm 0.18$ & $0.94 \pm 0.11$ &
 	\nodata & \nodata
\\
$[{\rm O\;II}]\;7320$ & $-0.400$ &
	$4.61 \pm 0.10\,$\tablenotemark{c} & $2.97 \pm 0.18$ &
	$5.98 \pm 0.27$ & $3.28\pm 0.21$ &
	\nodata & \nodata
\\
$[{\rm O\;II}]\;7330$ & $-0.402$ &
	$4.61 \pm 0.10\,$\tablenotemark{c} & $2.97 \pm 0.18$ &
	$4.43 \pm 0.25$ & $2.42 \pm 0.17$ &
	\nodata & \nodata
\\[1mm]
\multicolumn{2}{c}{$F(\hbeta)$ (ergs s$^{-1}$ cm$^{-2}$)} & 
        \multicolumn{2}{c}{$(3.15 \pm 0.12) \times 10^{-14}$} &
        \multicolumn{2}{c}{$(6.31 \pm 0.17) \times 10^{-15}$} &
        \multicolumn{2}{c}{$(1.90 \pm 0.14) \times 10^{-16}$} 
\\
\multicolumn{2}{c}{EW$_{\rm e}$(\hbeta) (\AA)} &
        \multicolumn{2}{c}{$288 \pm 58$} &
        \multicolumn{2}{c}{$207 \pm 25$} &
        \multicolumn{2}{c}{\nodata\tablenotemark{d}} 
\\
\multicolumn{2}{c}{Derived $E(B-V)$ (mag)$\,$\tablenotemark{e}} &
        \multicolumn{2}{c}{$+0.34 \pm 0.10$} &
        \multicolumn{2}{c}{$+0.466 \pm 0.081$} &
        \multicolumn{2}{c}{$+0.26 \pm 0.13$} 
\\
\multicolumn{2}{c}{$c(\hbeta)$} &
        \multicolumn{2}{c}{$0.477 \pm 0.061$} &
        \multicolumn{2}{c}{$0.653 \pm 0.049$} &
        \multicolumn{2}{c}{\nodata} 
\\
\multicolumn{2}{c}{Adopted $A_V$ (mag)} &
        \multicolumn{2}{c}{$+1.01$} &
        \multicolumn{2}{c}{$+1.39$} &
        \multicolumn{2}{c}{$+0.80$} 
\\
\multicolumn{2}{c}{EW$_{\rm abs}$ (\AA)} &
        \multicolumn{2}{c}{0} &
        \multicolumn{2}{c}{0} &
        \multicolumn{2}{c}{2} 
\\[1mm]
\tableline
& & \multicolumn{2}{c}{KD 9} & 
\multicolumn{2}{c}{KD 20} &    
\multicolumn{2}{c}{KD 21} \\   
\multicolumn{1}{c}{Property} &
\multicolumn{1}{c}{$f(\lambda)$} &
\multicolumn{1}{c}{$F$} & \multicolumn{1}{c}{$I$} &
\multicolumn{2}{c}{$F$} & 
\multicolumn{1}{c}{$F$} & \multicolumn{1}{c}{$I$} \\
\tableline
$[\rm{O\;II}]\;3727$ & $+0.325$ &
	$311 \pm 29$ & $377 \pm 35$ &
	\multicolumn{2}{c}{$789 \pm 65$} &
	$335.8 \pm 7.4$ & $427 \pm 25$ 
\\
${\rm H}\delta\;4101$ & $+0.232$ &
	\nodata & \nodata &
	\multicolumn{2}{c}{\nodata} &
	$17.2 \pm 2.0$ & $25.1 \pm 2.6$
\\
${\rm H}\gamma\;4340$ & $+0.158$ &
	\nodata & \nodata &
	\multicolumn{2}{c}{\nodata} &
	$44.2 \pm 2.8$ & $52.4 \pm 3.4$
\\
${\rm H}\beta\;4861$ & \phs0.000 &
	$100 \pm 14$ & $100 \pm 14$ &
	\multicolumn{2}{c}{$100 \pm 23$} &
	$100.0 \pm 4.2$ & $100.0 \pm 4.1$
\\
$[{\rm O\;III}]\;4959$ & $-0.026$ &
	$50 \pm 11$ & $48 \pm 11$ &
	\multicolumn{2}{c}{\nodata} &
	$43.0 \pm 3.4$ & $40.8 \pm 3.2$
\\
$[{\rm O\;III}]\;5007$ & $-0.038$ &
	$87 \pm 14$ & $83 \pm 13$ &
	\multicolumn{2}{c}{\nodata} &
	$112.8 \pm 4.2$ & $106.0 \pm 4.0$
\\
${\rm He\;I}\;5876$ & $-0.204$ &
	\nodata & \nodata &
	\multicolumn{2}{c}{\nodata} &
	$10.6 \pm 1.7$ & $8.7 \pm 1.4$
\\
${\rm H}\alpha\;6563$ & $-0.299$ &
	$353 \pm 23$ & $286 \pm 18$ &
	\multicolumn{2}{c}{$536 \pm 34$} &
	$393.0 \pm 7.9$ & $298 \pm 16$
\\
$[{\rm N\;II}]\;6583$ & $-0.301$ &
	$10 \pm 20$ & $8 \pm 16$ &
	\multicolumn{2}{c}{$10 \pm 28$} &
	$10.1 \pm 6.5$ & $7.6 \pm 4.9$
\\
$[{\rm S\;II}]\;6716$ & $-0.319$ &
	$27 \pm 17$ & $22 \pm 13$ &
	\multicolumn{2}{c}{$226 \pm 28\,$\tablenotemark{b}} &
	$48.4 \pm 5.9$ & $35.9 \pm 4.8$
\\
$[{\rm S\;II}]\;6731$ & $-0.321$ &
	$36 \pm 17$ & $28 \pm 13$ &
	\multicolumn{2}{c}{$226 \pm 28\,$\tablenotemark{b}} &
	$28.6 \pm 6.0$ & $21.2 \pm 4.6$
\\[1mm]
\multicolumn{2}{c}{$F(\hbeta)$ (ergs s$^{-1}$ cm$^{-2}$)} & 
        \multicolumn{2}{c}{$(5.57 \pm 0.77) \times 10^{-17}$} &
        \multicolumn{2}{c}{$(5.2 \pm 1.2) \times 10^{-17}$} &
        \multicolumn{2}{c}{$(6.40 \pm 0.27) \times 10^{-16}$} 
\\
\multicolumn{2}{c}{EW$_{\rm e}$(\hbeta) (\AA)} &
        \multicolumn{2}{c}{$78 \pm 18$} &
        \multicolumn{2}{c}{\nodata\tablenotemark{d}} &
        \multicolumn{2}{c}{$65.0 \pm 4.0$} 
\\
\multicolumn{2}{c}{Derived $E(B-V)$ (mag)$\,$\tablenotemark{e}} &
        \multicolumn{2}{c}{$+0.20 \pm 0.23$} &
        \multicolumn{2}{c}{$+0.65 \pm 0.33$} &
        \multicolumn{2}{c}{$+0.295 \pm 0.092$} 
\\
\multicolumn{2}{c}{$c(\hbeta)$} &
        \multicolumn{2}{c}{0.29} &
        \multicolumn{2}{c}{\nodata} &
        \multicolumn{2}{c}{$0.363 \pm 0.072$} 
\\
\multicolumn{2}{c}{Adopted $A_V$ (mag)} &
        \multicolumn{2}{c}{$+0.62$} &
        \multicolumn{2}{c}{$+2.0$} &
        \multicolumn{2}{c}{$+0.77$} 
\\
\multicolumn{2}{c}{EW$_{\rm abs}$ (\AA)} &
        \multicolumn{2}{c}{2} &
        \multicolumn{2}{c}{2} &
        \multicolumn{2}{c}{2}
\\[1mm]
\tableline
\end{tabular}
\vspace*{1mm}
\tablenotetext{a}{
Blended with $[{\rm Ne\;III}]\lambda\;3967$.
}
\tablenotetext{b}{
[S~II] unresolved.
}
\tablenotetext{c}{
\otwored\ unresolved.
}
\tablenotetext{d}{
Very weak continuum.
}
\tablenotetext{e}{
Derived from $F$(\halpha)/$F$(\hbeta).
}
\tablecomments{
See Table~\ref{table_gr07data1} for additional comments.
}
\end{center}
\end{table}

\begin{table}
\setlength{\tabcolsep}{2mm}
\tablenum{5c}
\begin{center}
\renewcommand{\arraystretch}{0.90} 
\caption{
Low-dispersion spectra: line ratios and properties for nebulae
KD~24, KD~25, KD~28, and KD~28e.
\vspace*{3mm}
\label{table_gr11data3}
}
\begin{tabular}{rccccccc}
\tableline \tableline
& & \multicolumn{2}{c}{KD 24} & 
\multicolumn{2}{c}{KD 25} &     
\multicolumn{2}{c}{KD 28} \\	
\multicolumn{1}{c}{Property} &
\multicolumn{1}{c}{$f(\lambda)$} &
\multicolumn{1}{c}{$F$} & \multicolumn{1}{c}{$I$} &
\multicolumn{1}{c}{$F$} & \multicolumn{1}{c}{$I$} &
\multicolumn{1}{c}{$F$} & \multicolumn{1}{c}{$I$} \\
\tableline
$[\rm{O\;II}]\;3727$ & $+0.325$ &
	$377 \pm 14$ & $813 \pm 90$ &
	$371 \pm 11$ & $539 \pm 17$ &
	$315 \pm 19$ & $468 \pm 28$ 
\\
${\rm H}\gamma\;4340$ & $+0.158$ &
	$42.6 \pm 4.4$ & $55.8 \pm 9.8$ &
	$32.7 \pm 6.3$ & $42.1 \pm 7.5$ &
	$40.9 \pm 7.4$ & $52.6 \pm 8.9$
\\
${\rm H}\beta\;4861$ & \phs0.000 &
	$100.0 \pm 4.5$ & $100.0 \pm 6.2$ &
	$100.0 \pm 5.4$ & $100.0 \pm 5.3$ &
	$100.0 \pm 8.8$ & $100.0 \pm 8.6$
\\
$[{\rm O\;III}]\;4959$ & $-0.026$ &
	$7.2 \pm 3.3$ & $6.9 \pm 3.6$ &
	$16.8 \pm 4.1$ & $15.9 \pm 3.9$ &
	$19.9 \pm 7.1$ & $18.8 \pm 6.7$
\\
$[{\rm O\;III}]\;5007$ & $-0.038$ &
	$19.3 \pm 3.5$ & $18.1 \pm 4.5$ &
	$35.8 \pm 4.5$ & $33.3 \pm 4.2$ &
	$78.6 \pm 8.5$ & $73.2 \pm 7.9$
\\
${\rm H}\alpha\;6563$ & $-0.299$ &
	$496 \pm 11$ & $286 \pm 29$ &
	$416.9 \pm 9.3$ & $283.9 \pm 6.3$ &
	$432 \pm 19$ & $288 \pm 12$
\\
$[{\rm N\;II}]\;6583$ & $-0.301$ &
	$26.1 \pm 9.6$ & $15.0 \pm 6.5$ &
	$9.5 \pm 7.8$ & $6.4 \pm 5.3$ &
	$3 \pm 15$ & $3 \pm 12$
\\
$[{\rm S\;II}]\;6716, 6731$ & $-0.320$ &
	$120.9 \pm 8.4$ & $67.5 \pm 9.4$ &
	$72.3 \pm 6.7$ & $47.7 \pm 4.4$ &
	$77 \pm 13$ & $49.9 \pm 8.7$
\\[1mm]
\multicolumn{2}{c}{$F(\hbeta)$ (ergs s$^{-1}$ cm$^{-2}$)} & 
        \multicolumn{2}{c}{$(6.77 \pm 0.30) \times 10^{-16}$} &
        \multicolumn{2}{c}{$(1.243 \pm 0.067) \times 10^{-16}$} &
        \multicolumn{2}{c}{$(4.89 \pm 0.43) \times 10^{-17}$} 
\\
\multicolumn{2}{c}{EW$_{\rm e}$(\hbeta) (\AA)} &
        \multicolumn{2}{c}{\nodata\tablenotemark{a}} &
        \multicolumn{2}{c}{$81.2 \pm 7.0$} &
        \multicolumn{2}{c}{$93 \pm 15$} 
\\
\multicolumn{2}{c}{Derived $E(B-V)$ (mag)$\,$\tablenotemark{b}} &
        \multicolumn{2}{c}{$+0.56 \pm 0.10$} &
        \multicolumn{2}{c}{$+0.36 \pm 0.10$} &
        \multicolumn{2}{c}{$+0.40 \pm 0.15$} 
\\
\multicolumn{2}{c}{$c(\hbeta)$} &
        \multicolumn{2}{c}{\nodata} &
        \multicolumn{2}{c}{0.53} &
        \multicolumn{2}{c}{0.56} 
\\
\multicolumn{2}{c}{Adopted $A_V$ (mag)} &
        \multicolumn{2}{c}{$+1.72$} &
        \multicolumn{2}{c}{$+1.13$} &
        \multicolumn{2}{c}{$+1.19$} 
\\
\multicolumn{2}{c}{EW$_{\rm abs}$ (\AA)} &
        \multicolumn{2}{c}{2} &
        \multicolumn{2}{c}{2} &
        \multicolumn{2}{c}{2} 
\\[1mm]
\tableline
& & \multicolumn{2}{c}{KD 28e} & 
\\
\multicolumn{1}{c}{Property} &
\multicolumn{1}{c}{$f(\lambda)$} &
\multicolumn{1}{c}{$F$} & \multicolumn{1}{c}{$I$} &
\\
\tableline
$[\rm{O\;II}]\;3727$ & $+0.325$ &
	$229.2 \pm 3.1$ & $305.3 \pm 9.6$ &
\\
${\rm H}11\;3772$ & $+0.316$ &
	$3.4 \pm 2.4$ & $4.5 \pm 3.2$ &
\\
${\rm H}10\;3799$ & $+0.310$ &
	$4.8 \pm 2.6$ & $6.3 \pm 3.4$ &
\\
${\rm H}9\;3835$ & $+0.302$ &
	$6.41 \pm 0.60$ & $8.36 \pm 0.81$ &
\\
$[{\rm Ne\;III}]\;3869$ & $+0.294$ &
	$27.46 \pm 0.62$ & $35.6 \pm 1.2$ &
\\
${\rm H}8 + {\rm He\;I}\;3889$ & $+0.289$ &
	$17.96 \pm 0.59$ & $23.17 \pm 0.96$ &
\\
${\rm H}\epsilon + {\rm He\;I}\;3970\,$\tablenotemark{c} & $+0.269$ &
	$20.96 \pm 0.62$ & $26.6 \pm 1.0$ &
\\
$[{\rm S\;II}]\;4068,4076$ & $+0.241$ &
	$1.45 \pm 0.46$ & $1.79 \pm 0.57$ &
\\
${\rm H}\delta\;4101$ & $+0.232$ &
	$22.67 \pm 0.59$ & $27.81 \pm 0.92$ &
\\
${\rm H}\gamma\;4340$ & $+0.158$ &
	$42.42 \pm 0.66$ & $48.8 \pm 1.0$ &
\\
$[{\rm O\;III}]\;4363$ & $+0.151$ &
	$3.49 \pm 0.52$ & $3.99 \pm 0.60$ &
\\
${\rm He\;I}\;4471$ & $+0.116$ &
	$3.86 \pm 0.34$ & $4.28 \pm 0.38$ &
\\
${\rm H}\beta\;4861$ & \phs0.000 &
	$100.0 \pm 2.5$ & $100.0 \pm 2.5$ &
\\
$[{\rm O\;III}]\;4959$ & $-0.026$ &
	$138.0 \pm 6.3$ & $134.9 \pm 6.2$ &
\\
$[{\rm O\;III}]\;5007$ & $-0.038$ &
	$412.4 \pm 8.1$ & $398.7 \pm 7.9$ &
\\
${\rm He\;I}\;5876$ & $-0.204$ &
	$12.93 \pm 0.47$ & $10.80 \pm 0.44$ &
\\
$[{\rm O\;I}]\;6300 + [{\rm S\;III}]\;6312$ & $-0.264$ &
	$3.75 \pm 0.46$ & $2.97 \pm 0.37$ &
\\
$[{\rm O\;I}]\;6363$ & $-0.272$ &
	$1.14 \pm 0.36$ & $0.90 \pm 0.28$ &
\\
${\rm H}\alpha\;6563$ & $-0.299$ &
	$377.7 \pm 4.9$ & $290.1 \pm 8.5$ &
\\
$[{\rm N\;II}]\;6583$ & $-0.301$ &
        $9.4 \pm 1.4$ & $7.2 \pm 1.1$ &     
\\
${\rm He\;I}\;6678$ & $-0.314$ &
	$3.8 \pm 3.7$ & $2.9 \pm 2.8$ &
\\
$[{\rm S\;II}]\;6716$ & $-0.319$ &
	$16.7 \pm 3.8$ & $12.6 \pm 2.9$ &
\\
$[{\rm S\;II}]\;6731$ & $-0.321$ &
	$12.9 \pm 3.9$ & $9.7 \pm 3.0$ &
\\
${\rm He\;I}\;7065$ & $-0.366$ &
	$4.18 \pm 0.42$ & $3.03 \pm 0.32$ &
\\
$[{\rm Ar\;III}]\;7136$ & $-0.375$ &
	$17.33 \pm 0.49$ & $12.44 \pm 0.54$ &
\\
${\rm He\;I}\;7281$ & $-0.395$ &
	$1.46 \pm 0.40$ & $1.03 \pm 0.28$ &
\\
$[{\rm O\;II}]\;7320$ & $-0.400$ &
	$6.63 \pm 0.49$ & $4.66 \pm 0.38$ &
\\
$[{\rm O\;II}]\;7330$ & $-0.401$ &
	$3.62 \pm 0.49$ & $2.54 \pm 0.36$ &
\\[1mm]
\multicolumn{2}{c}{$F(\hbeta)$ (ergs s$^{-1}$ cm$^{-2}$)} & 
        \multicolumn{2}{c}{$(3.782 \pm 0.094) \times 10^{-15}$} &
\\
\multicolumn{2}{c}{EW$_{\rm e}$(\hbeta) (\AA)} &
        \multicolumn{2}{c}{$337 \pm 56$} &
\\
\multicolumn{2}{c}{Derived $E(B-V)$ (mag)$\,$\tablenotemark{b}} &
        \multicolumn{2}{c}{$+0.277 \pm 0.074$} &
\\
\multicolumn{2}{c}{$c(\hbeta)$} &
        \multicolumn{2}{c}{$0.383 \pm 0.039$} &
\\
\multicolumn{2}{c}{Adopted $A_V$ (mag)} &
        \multicolumn{2}{c}{$+0.81$} &
\\
\multicolumn{2}{c}{EW$_{\rm abs}$ (\AA)} &
        \multicolumn{2}{c}{0} &
\\[1mm]
\tableline
\end{tabular}
\vspace*{2mm}
\tablenotetext{a}{
Very weak continuum.
}
\tablenotetext{b}{
Derived from $F$(\halpha)/$F$(\hbeta).
}
\tablenotetext{c}{
Blended with [\ion{Ne}{3}]$\lambda$ 3967.
}
\tablecomments{
See Table~\ref{table_gr07data1} for additional comments.
}
\end{center}
\end{table}

\clearpage

\begin{table}
\setlength{\tabcolsep}{1.2mm}
\tablenum{6a}
\begin{center}
\renewcommand{\arraystretch}{1.1}
\caption{
Ionic and total abundances.
\label{table_abund1}
}
%
\begin{tabular}{lcccccccccc}
\tableline \tableline
& & & & & & & & Hu I & Hu I & Hu III
\\
& HK 16 & HK 42 & HK 69 & HK 70 & Ho 7 & Ho 12 & Ho 12 & 
E ctr\tablenotemark{a} & NW spot\tablenotemark{a} & 
SE ring\tablenotemark{a} 
\\
Property & Low disp. & Low disp. & Low disp. & Low disp. & 
Low disp. & Blue & Low disp. & Blue & Blue & Blue 
\\
\tableline
$T_e$(O$^{+2}$) (K) &
\nodata & \nodata & \nodata & \nodata & \nodata & \nodata & \nodata & 
\nodata & $13100 \pm 1500$ & \nodata 
\\
$T_e$(O$^+$) (K)$\,$\tablenotemark{b} &
\nodata & \nodata & \nodata & \nodata & \nodata & \nodata & \nodata & 
\nodata & $12200 \pm 1400$ & \nodata 
\\
O$^+$/H $(\times 10^5)$ & 
\nodata & \nodata & \nodata & \nodata & \nodata & \nodata & \nodata & 
\nodata & $3.9 \pm 1.6$ & \nodata
\\
O$^{+2}$/H $(\times 10^5)$ &
\nodata & \nodata & \nodata & \nodata & \nodata & \nodata & \nodata &
\nodata & $5.5 \pm 1.6$ & \nodata 
\\
O/H $(\times 10^5)$ & 
\nodata & \nodata & \nodata & \nodata & \nodata & \nodata & \nodata &
\nodata & $9.4 \pm 2.3$ & \nodata 
\\
12$+$log(O/H) &
\nodata & \nodata & $> 8.10\,$\tablenotemark{c} & 
\nodata & \nodata & \nodata & \nodata &
\nodata & $7.98 \pm 0.09 \, (^{+0.11}_{-0.15})$ & \nodata 
\\
12$+$log(O/H) M91$\,$\tablenotemark{d} &
8.28 & \nodata\tablenotemark{e} & \nodata & 
\nodata\tablenotemark{e} &
8.26 & 8.18 & 8.32 & 8.05 & 8.05 & 8.16
\\
12$+$log(O/H) P00$\,$\tablenotemark{f} &
\nodata & 8.43 & \nodata & 8.30 & 
8.38 & \nodata & \nodata & 7.88 & 7.89 & 8.09
\\[1mm]
\tableline
Ne$^{+2}$/O$^{+2}$ &
\nodata & \nodata & \nodata & \nodata & \nodata & \nodata & \nodata &
\nodata & \phs$0.23 \pm 0.10$ & \nodata 
\\
log(Ne/O) &
\nodata & \nodata & \nodata & \nodata & \nodata & \nodata & \nodata &
\nodata & $-0.64 \pm 0.19$ & \nodata 
\\
\tableline
\end{tabular}
\vspace*{2mm}
\tablenotetext{a}{
See Fig.~\ref{fig_n6822_hu3} for locations of resolved features.
}
\tablenotetext{b}{
O$^+$ zone temperature derived using Equation~\ref{eqn_toplus};
see \cite{ctm86} and \cite{garnett92}.
}
\tablenotetext{c}{\scriptsize
Lower limit to the oxygen abundance, as \othree\ was not detected.
}
\tablenotetext{d}{
\cite{mcgaugh91} bright-line calibration.
}
\tablenotetext{e}{
$\log \, R_{23} \, \ga \, 1$.
}
\tablenotetext{f}{
\cite{pilyugin00} bright-line calibration.
}
\tablecomments{
Nebulae are listed in the same order as they appear in
Table~\ref{table_obslog}.
Direct oxygen abundances are shown with two uncertainties.
The first uncertainty is the formal uncertainty in the derivation.
In parentheses is the range of possible values, expressed by the
maximum and minimum values of the oxygen abundance.
}
\end{center}
\end{table}

\begin{table}
\scriptsize  
\setlength{\tabcolsep}{1.mm}
\tablenum{6b}
\begin{center}
\renewcommand{\arraystretch}{0.9}
\caption{
Ionic and total abundances (continued).
\label{table_abund2}
}
%
\begin{tabular}{lcccccccc}
\tableline \tableline
& Hu III & Hu III
\\
& 
NW arc\tablenotemark{a} & NW ring\tablenotemark{a} &
Hu V & Hu V & K$\alpha$ & K$\alpha$ & K$\beta$ & K$\beta$
\\
Property & Blue & Blue & Blue & Low disp. & Blue & Low disp. & 
Blue & Low disp.
\\
\tableline
$T_e$(O$^{+2}$) (K) &
\nodata & \nodata & 
$11930 \pm 500$ & $11860 \pm 640$ &  
$12660 \pm 590$ & $14060 \pm 660$ &  
$11840 \pm 900$ & \nodata   
\\
$T_e$(O$^+$) (K)$\,$\tablenotemark{b} &
\nodata & \nodata & 
$11350 \pm 500$ & $11300 \pm 610$ &  
$11860 \pm 560$ & $12840 \pm 610$ &  
$11290 \pm 860$ & \nodata   
\\
$T_e$(O$^+$) (K)$\,$\tablenotemark{c} &
\nodata & \nodata & 
\nodata & $11200 \pm 750$ &  
\nodata & $15900 \pm 1600$ &  
\nodata & \nodata   
\\
$T_e$(S$^+$) (K)$\,$\tablenotemark{d} &
\nodata & \nodata & 
\nodata & $13000 \pm 10000$ &  
\nodata & $9800 \pm $ 9200&  
\nodata & \nodata   
\\
O$^+$/H $(\times 10^5)$ & 
\nodata & \nodata & 
$3.11 \pm 0.52$ & $2.90 \pm 0.61$ &  
$1.83 \pm 0.31$ & $1.76 \pm 0.29$ &  
$4.6 \pm 1.3$ & \nodata   
\\
O$^{+2}$/H $(\times 10^5)$ &
\nodata & \nodata & 
$10.3 \pm 1.2$ & $11.1 \pm 1.7$ &  
$9.6 \pm 1.2$ & $7.06 \pm 0.83$ &  
$10.4 \pm 2.2$ & \nodata   
\\
O/H $(\times 10^5)$ & 
\nodata & \nodata & 
$13.4 \pm 1.3$ & $13.9 \pm 1.8$ &  
$11.4 \pm 1.2$ & $8.83 \pm 0.88$ &  
$15.0 \pm 2.6$ & \nodata   
\\
12$+$log(O/H) &
\nodata & \nodata & 
$8.13 \pm 0.04\,(^{+0.05}_{-0.06})$ & 
$8.15 \pm 0.05\,(^{+0.06}_{-0.07})$ & 
$8.06 \pm 0.04\,(^{+0.05}_{-0.06})$ & 
$7.95 \pm 0.04\,(^{+0.05}_{-0.06})$ & 
$8.18 \pm 0.07\,(^{+0.09}_{-0.11})$ &  
\nodata  
\\
12$+$log(O/H) M91$\,$\tablenotemark{e} &
8.09 & 7.99 & 8.06 & 8.07 & 8.03 & 8.07 & 8.16 & 8.37 
\\
12$+$log(O/H) P00$\,$\tablenotemark{f} &
8.18 & 8.13 & 7.83 & 7.83 & 7.80 & 7.83 & 7.92 & \nodata
\\[1mm]
\tableline
Ar$^{+2}$/H $(\times 10^7)$ &
\nodata & \nodata & \nodata &
$6.48 \pm 0.98$ & 
\nodata & $5.28 \pm 0.29$ &  
\nodata & \nodata   
\\
Ar$^{+3}$/H $(\times 10^7)$ &
\nodata & \nodata & \nodata &
$0.29 \pm 0.13$ &  
\nodata & \nodata &  
\nodata & \nodata  
\\
ICF(Ar) &
\nodata & \nodata & \nodata &
1.05 &  
\nodata & 1.80 &  
\nodata & \nodata   
\\
Ar/H $(\times 10^7)$ &
\nodata & \nodata & \nodata &
$7.1 \pm 1.2$ &  
\nodata & $9.5 \pm 1.1$ &  
\nodata & \nodata  
\\
log(Ar/O) &
\nodata & \nodata & \nodata &
$-2.29 \pm 0.04$ &  
\nodata & $-1.96 \pm 0.03$ &  
\nodata & \nodata   
\\
N$^+$/O$^+$ ($\times 10^2)$ &
\nodata & \nodata & \nodata &
$0.703 \pm 0.067$ & 
\nodata & $1.44 \pm 0.20$ & 
\nodata & \nodata
\\
log(N/O) &
\nodata & \nodata & \nodata &
$-2.15 \pm 0.04$ & 
\nodata & $-1.84 \pm 0.06$ & 
\nodata & \nodata
\\
Ne$^{+2}$/O$^{+2}$ &
\nodata & \nodata & 
\phs$0.208 \pm 0.053$ & \phs$0.267 \pm 0.062$ &  
\phs$0.245 \pm 0.063$ & \phs$0.324 \pm 0.067$ &  
\phs$0.261 \pm 0.080$ & \nodata   
\\
log(Ne/O) &
\nodata & \nodata & 
$-0.68 \pm 0.11$ & $-0.57 \pm 0.10$ &  
$-0.61 \pm 0.11$ & $-0.489 \pm 0.091$ &  
$-0.58 \pm 0.13$ & \nodata  
\\
\tableline
\end{tabular}
\vspace*{2mm}
\tablenotetext{a}{
See Fig.~\ref{fig_n6822_hu3} for locations of resolved features.
}
\tablenotetext{b}{
O$^+$ zone temperature derived using Equation~\ref{eqn_toplus};
see \cite{ctm86} and \cite{garnett92}.
}
\tablenotetext{c}{
O$^+$ zone temperature derived from $I$(\otwo)/$I$(\otwored).
}
\tablenotetext{d}{
S$^+$ zone temperature derived from $I$(\stwo)/$I$(\stwoblue).
}
\tablenotetext{e}{
\cite{mcgaugh91} bright-line calibration.
}
\tablenotetext{f}{
\cite{pilyugin00} bright-line calibration.
}
\tablecomments{
See Table~\ref{table_abund1} for additional comments.
}
\end{center}
\end{table}

\begin{table}
\setlength{\tabcolsep}{1.2mm}
\tablenum{6c}
\begin{center}
\renewcommand{\arraystretch}{1.}
\caption{
Ionic and total abundances (continued).
\label{table_abund3}
}
%
\begin{tabular}{lccccccccc}
\tableline \tableline
& KD 9 & KD 20 & KD 21 & KD 22e & KD 24 & KD 25 & KD 28 & KD 28e & KD 28e 
\\
Property & Low disp. & Low disp. & Low disp. & Blue & Low disp. & 
Low disp. & Low disp. & Blue & Low disp.
\\
\tableline
$T_e$(O$^{+2}$) (K) &
\nodata & \nodata & \nodata & \nodata & \nodata & \nodata & \nodata &
$11530 \pm 610$ & $11540 \pm 620$
\\
$T_e$(O$^+$) (K)$\,$\tablenotemark{a} & 
\nodata & \nodata & \nodata & \nodata & \nodata & \nodata & \nodata &
$11070 \pm 580$ & $11080 \pm 600$
\\
$T_e$(O$^+$) (K)$\,$\tablenotemark{b} & 
\nodata & \nodata & \nodata & \nodata & \nodata & \nodata & \nodata &
\nodata & $11180 \pm 720$
\\
$T_e$(S$^+$) (K)$\,$\tablenotemark{c} & 
\nodata & \nodata & \nodata & \nodata & \nodata & \nodata & \nodata &
\nodata & $9500 \pm 4100$
\\
O$^+$/H $(\times 10^5)$ & 
\nodata & \nodata & \nodata & \nodata & \nodata & \nodata & \nodata &
$5.8 \pm 1.2$ & $7.4 \pm 1.6$
\\
O$^{+2}$/H $(\times 10^5)$ &
\nodata & \nodata & \nodata & \nodata & \nodata & \nodata & \nodata &
$9.2 \pm 1.4$ & $8.7 \pm 1.4$
\\
O/H $(\times 10^5)$ & 
\nodata & \nodata & \nodata & \nodata & $> 8.13\,$\tablenotemark{d} & 
\nodata & \nodata &
$15.0 \pm 1.8$ & $16.2 \pm 2.1$
\\
12$+$log(O/H) &
\nodata & \nodata & \nodata & \nodata & \nodata & \nodata & \nodata &
$8.18 \pm 0.05\,(^{+0.06}_{-0.07})$ & 
$8.21 \pm 0.05\,(^{+0.06}_{-0.07})$
\\
12$+$log(O/H) M91$\,$\tablenotemark{e} &
8.06 & \nodata & 8.15 & 8.22 & \nodata & 8.37 & 8.22 & 
8.12 & 8.20
\\
12$+$log(O/H) P00$\,$\tablenotemark{f} &
8.40 & \nodata & 8.48 & \nodata & \nodata & \nodata & \nodata & 
7.93 & 8.03 
\\
\tableline
Ar$^{+2}$/H $(\times 10^7)$ &
\nodata & \nodata & \nodata & \nodata & \nodata & \nodata & \nodata &
\nodata & $8.6 \pm 1.1$
\\
ICF(Ar) &
\nodata & \nodata & \nodata & \nodata & \nodata & \nodata & \nodata &
\nodata & 1.45
\\
Ar/H $(\times 10^7)$ &
\nodata & \nodata & \nodata & \nodata & \nodata & \nodata & \nodata &
\nodata & $12.5 \pm 1.6$
\\
log(Ar/O) &
\nodata & \nodata & \nodata & \nodata & \nodata & \nodata & \nodata &
\nodata & $-2.11 \pm 0.04$
\\
N$^+$/O$^+$ ($\times 10^2)$ &
\nodata & \nodata & \nodata & \nodata & \nodata & \nodata & \nodata &
\nodata & $1.46 \pm 0.23$
\\
log(N/O) &
\nodata & \nodata & \nodata & \nodata & \nodata & \nodata & \nodata &
\nodata & $-1.83 \pm 0.06$
\\
Ne$^{+2}$/O$^{+2}$ &
\nodata & \nodata & \nodata & \nodata & \nodata & \nodata & \nodata &
$0.204 \pm 0.053$ & $0.271 \pm 0.053$
\\
log(Ne/O) &
\nodata & \nodata & \nodata & \nodata & \nodata & \nodata & \nodata &
$-0.68 \pm 0.11$ & $-0.567 \pm 0.086$
\\
\tableline
\end{tabular}
\vspace*{2mm}
\tablenotetext{a}{
O$^+$ zone temperature derived using Equation~\ref{eqn_toplus};
see \cite{ctm86} and \cite{garnett92}.
}
\tablenotetext{b}{
O$^+$ zone temperature derived from $I$(\otwo)/$I$(\otwored).
}
\tablenotetext{c}{
S$^+$ zone temperature derived from $I$(\stwo)/$I$(\stwoblue).
}
\tablenotetext{d}{\scriptsize
Lower limit to the oxygen abundance, as \othree\ was barely
detected. 
}
\tablenotetext{e}{
\cite{mcgaugh91} bright-line calibration.
}
\tablenotetext{f}{
\cite{pilyugin00} bright-line calibration.
}
\tablecomments{
See Table~\ref{table_abund1} for additional comments.
}
\end{center}
\end{table}

\begin{table}
\small 
\setlength{\tabcolsep}{2mm}
\tablenum{7}
\begin{center}
\renewcommand{\arraystretch}{1.1}
\caption{
Oxygen abundances in NGC~6822.
\vspace*{3mm}
\label{table_alloxy}
}
\begin{tabular}{cccc}
\tableline \tableline
& Radius & Adopted & \\
Object & (kpc) & 12$+$log(O/H) & References \\
(1) & (2) & (3) & (4) \\
\tableline
\multicolumn{4}{c}{\sf \hii\ regions} \\
\tableline
HK 16 & 0.84 & $8.28 \pm 0.20$ & 1 \\
HK 42 & 0.90 & $8.43 \pm 0.20$ & 1 \\
HK 69 & 0.10 & $> 8.10\,$\tablenotemark{a} & 1 \\
HK 70 & 0.22 & $8.30 \pm 0.20$ & 1 \\
Hodge 5 & 1.11 & $8.23 \pm 0.20$ & 2 \\ 
Hodge 6 & 0.78 & $8.24 \pm 0.20$ & 2 \\
Hodge 7 & 0.45 & $8.26 \pm 0.20$ & 1 \\
Hodge 8 & 1.11 & $8.26 \pm 0.15$ & 3 \\
Hodge 10$\,$\tablenotemark{b} & 0.91 & 
	$\mathbf{8.07 \pm 0.08}$ & 3, {\bf 4} \\  
Hodge 12$\,$\tablenotemark{c} & 0.08 & 
	$7.9 \pm 0.1$ & 1 \\
Hubble I & 1.37 & $\mathbf{7.98 \pm 0.09}$ & {\bf 1}, 2, 3 \\ 
Hubble II & 1.02 & $8.27 \pm 0.20$ & 2 \\
Hubble III & 1.28 & $8.11 \pm 0.20$ & 1, 2, 3 \\
Hubble V & 1.03 & $\mathbf{8.14 \pm 0.05}$ & 
	{\bf 1}, 2, 3, 5, 6, 7, 8, 9, 10, 11 \\
Hubble X & 1.20 & $\mathbf{8.01 \pm 0.05}$ & 
	2, 3, 6, 7, 9, 10, {\bf 11} \\
K$\alpha$ & 1.00 & $\mathbf{8.00 \pm 0.04}$ & {\bf 1}, 4 \\
K$\beta$ & 0.87 & $\mathbf{8.18 \pm 0.07}$ & {\bf 1}, 4 \\
K$\gamma$ & 1.08 & 
	$\mathbf{8.02 \pm 0.04}$ & {\bf 4} \\ 
KD 9 & 0.87 & $8.23 \pm 0.20$ & 1 \\
KD 12$\,$\tablenotemark{d} & 0.98 & $> 7.80$ & 4 \\  
KD 13 & 1.16 & $8.26 \pm 0.16$ & 3 \\
KD 20$\,$\tablenotemark{d} & 0.21 & $> 7.93$ & 1, 4 \\  
KD 21 & 1.06 & $8.31 \pm 0.20$ & 1  \\ 
KD 22e & 1.02 & $8.22 \pm 0.20$ & 1 \\
KD 24 & 0.17 & $> 8.13\,$\tablenotemark{a} & 1 \\ 
KD 25 & 0.40 & $8.37 \pm 0.20$ & 1 \\
KD 28 & 1.90 & $8.22 \pm 0.20$ & 1 \\
KD 28e & 0.07 & $\mathbf{8.19 \pm 0.05}$ & {\bf 1} \\
\tableline
\multicolumn{4}{c}{\sf Planetary nebulae} \\
\tableline
S16 & 0.25 & $8.01 \pm 0.16$ & 12 \\
S33 & 0.48 & $8.10 \pm 0.08$ & 12, 13 \\
\tableline
\multicolumn{4}{c}{\sf A-type supergiant stars} \\
\tableline
CW 22 & 1.23 & $8.20 \pm 0.10$ & 14 \\
CW 173 & 0.40 & $8.28 \pm 0.17$ & 15, 16 \\ 
CW 185 & 0.29 & $8.44 \pm 0.09$ & 15, 16 \\ 
\tableline
\end{tabular}
\vspace*{2mm}
\tablenotetext{a}{\small
Lower limits due to zero and noisy detection of [\ion{O}{3}] 
$\lambda5007$ in HK~69 and KD~24, respectively.
}
\tablenotetext{b}{\small
Also known as Hubble~IV, KD~18.
}
\tablenotetext{c}{\small
Supernova remnant.
}
\tablenotetext{d}{\small
KD~12 and KD~20 identified as C9 and C10, respectively, 
in \cite{chandar00}.
}
\tablecomments{
Col.~(1): Identifications - \hii\ regions from \cite{kd82} and
\cite{hkl88}; planetary nebulae from \cite{kd82};
A-type supergiant stars from \cite{wilson92}.
Col.~(2): Deprojected galactocentric radius, calculated using the
parameters listed in Table~\ref{table_n6822}.
Cols.~(3) and (4): Oxygen abundances and references, respectively.
}
\tablerefs{
(1) present work;
(2) \cite{pes80};
(3) \cite{talent80};
(4) \cite{chandar00};
(5) \cite{ps70};
(6) \cite{smith75};
(7) \cite{lequeux79};
(8) \cite{stm89};
(9) \cite{hgom01};
(10) \cite{miller96};
(11) \cite{peimbert05};
(12) \cite{rm95};
(13) \cite{dt80};
(14) \cite{lsv03};
(15) \cite{venn01};
(16) \cite{vm02}.
}
\end{center}
\end{table}

\begin{table}
\small  
\setlength{\tabcolsep}{2mm}
\tablenum{8}
\begin{center}
\renewcommand{\arraystretch}{1.1}
\caption{
Linear least-squares fits to various subsets of oxygen abundance data.
\vspace*{3mm}
\label{table_fits}
}
\begin{tabular}{lcc}
\tableline \tableline
& Linear & Rms about \\
Data subset & Least-Squares Fit & fit (dex) \\
(1) & (2) & (3) \\
\tableline
(a): All nebulae in present work & 
${\cal Y} = (8.21 \pm 0.07) + (-0.02 \pm 0.07)\,R$ & 0.14 \\
%
(b): A-type supergiants $+$ set (a) &
${\cal Y} = (8.25 \pm 0.06) + (-0.05 \pm 0.07)\,R$ & 0.14 \\
%
(c): \hii\ s with \othreea: present work &
${\cal Y} = (8.23 \pm 0.06) + (-0.14 \pm 0.07)\,R$ & 0.06 \\
%
(d): A-type supergiants &
${\cal Y} = (8.43 \pm 0.06) + (-0.19 \pm 0.08)\,R$ & 0.06 \\
%
(e): \hii\ s with \othreea: set (c) $+$ liter. &
${\cal Y} = (8.23 \pm 0.05) + (-0.16 \pm 0.05)\,R$ & 0.05 \\
%
(f): Sets (d) $+$ (e) &
${\cal Y} = (8.34 \pm 0.07) + (-0.24 \pm 0.07)\,R$ & 0.09 \\
%
\tableline
\end{tabular}
\vspace*{2mm}
\tablecomments{
${\cal Y}$ represents the oxygen abundance, 12$+$log(O/H), and
$R$ represents the linear galactocentric radius in kpc.
Dataset a is the set of measurements for all nebulae in the
present work.
Dataset b combines the previous dataset with measurements for A-type
supergiants.
Dataset c is the set of \othreea\ measurements for the five \hii\
regions in the present work.
Dataset d is a fit only to the A-type supergiants.
Dataset e is the set of \othreea\ measurements for the five \hii\
regions in this work and three other \othreea\ detections from the
literature (\citealp{chandar00,peimbert05}).
We adopt oxygen abundances for all \hii\ regions which were remeasured
in the present work (e.g., Hubble~V). 
Dataset f combines the previous dataset with measurements for A-type
supergiants.
The obtained linear least-squares fit to every dataset is shown 
as a solid line in the corresponding panel in Fig.~\ref{fig_gradient2}. 
}
\end{center}
\end{table}

\end{document}